\documentclass[useAMS,usenatbib]{mn2e}
\usepackage{ gensymb }
\usepackage[pdftex]{graphicx}
\usepackage[fleqn]{amsmath}
\usepackage{amssymb}
\usepackage{bm}
\usepackage{epstopdf}

\newcommand{\ie}{i.e. }

\newcommand{\dd}{{\rm d}}

\title[MRI with buoyancy]
{Numerical simulations of the magnetorotational instability in
  protoneutron stars : I. Influence of buoyancy}
  
  \author[Guilet \& M\"uller]{J\'er\^ome Guilet$^{1,2}$ \& Ewald M\"uller$^{1}$ \\
$^1$ Max-Planck-Institut f\"ur Astrophysik, Karl-Schwarzschild-Str. 1, D-85748 Garching, Germany \\ 
$^2$ Max Planck/Princeton Center for Plasma Physics}

\begin{document}

\maketitle

\label{firstpage}

\begin{abstract}
  The magneto-rotational instability (MRI) is considered to be a
  promising mechanism to amplify the magnetic field in fast rotating
  protoneutron stars. In contrast to accretion discs, radial buoyancy
  driven by entropy and lepton fraction gradients is expected to have
  a dynamical role as important as rotation and shear. We investigate
  the poorly known impact of buoyancy on the non-linear phase of the
  MRI, by means of three dimensional numerical simulations of a local
  model in the equatorial plane of a protoneutron star. The use of
  the Boussinesq approximation allows us to utilise a shearing box
  model with clean shearing periodic boundary conditions, while taking
  into account the buoyancy driven by radial entropy and composition
  gradients. We find significantly stronger turbulence and magnetic
  fields in buoyantly unstable flows. On the other hand, buoyancy has
  only a limited impact on the strength of turbulence and magnetic
  field amplification for buoyantly stable flows in the presence of a
  realistic thermal diffusion. The properties of the turbulence are,
  however, significantly affected in the latter case. In particular,
  the toroidal components of the magnetic field and of the velocity
  become even more dominant with respect to the poloidal
  ones. Furthermore, we observed in the regime of stable buoyancy the
  formation of long lived coherent structures such as channel flows
  and zonal flows. Overall, our results support the ability of the MRI
  to amplify the magnetic field significantly even in stably
  stratified regions of protoneutron stars.
\end{abstract}


\section{Introduction}
The impact of rotation and magnetic field on the dynamics of core
collapse supernovae is still uncertain and represents a promising
avenue of research. In the case of fast rotating progenitors giving
birth to a protoneutron star (PNS) rotating with a period of a few
milliseconds, the rotational energy represents an important energy
reservoir that can be efficiently tapped if strong magnetic fields are
present. In these conditions, the magnetorotational instability (MRI)
has been proposed as a mechanism to efficiently amplify the magnetic
field \citep{akiyama03}. Its development could impact the explosion by
converting shear energy into thermal energy \citep{thompson05} or by
generating a strong large-scale magnetic field which can extract the
rotational energy and launch a magnetorotational explosion
\citep{leblanc70, bisnovatyi-Kogan76, mueller79, symbalisty84,
  moiseenko06, shibata06,burrows07b, dessart08, takiwaki09,
  takiwaki11, winteler12, mosta14}. The magnetic field amplification
by the MRI to a strength of $10^{15}\,{\rm G}$ could furthermore be a
scenario to explain the formation of the most strongly magnetized
neutron stars known as magnetars \citep[][and references
therein]{woods06}. The birth of such magnetars with rotation periods
in the millisecond range is a potential central engine for long
gamma-ray bursts \citep[e.g.][]{duncan92,metzger11} associated with
supernovae that have extreme kinetic energies \citep[also called
"hypernovae" or type Ic BL, e.g.][]{drout11}. The delayed energy
injection due to the spin down of a fast rotating, highly magnetized
neutron star has also been invoked as an explanation for some
superluminous supernovae like SN 2008 bi \citep{kasen10, woosley10,
  dessart12, nicholl13, inserra13}.

Since the seminal work of \citet{balbus91}, the MRI has been thoroughly investigated in the context of
accretion discs. The physical conditions in the PNS differ from those
prevailing in accretion discs in several aspects. The main support
against gravity is provided by pressure rather than centrifugal
forces, and as a consequence the rotation profile can be
non-Keplerian. A strong differential rotation appropriate for MRI growth nonetheless results from the collapse of the progenitor core to a PNS \citep{akiyama03,ott06}. The dependence of MRI turbulence on the steepness of
the rotation profile was studied by \citet{masada12}. Another
consequence of the high pressure is that radial buoyancy forces driven
by entropy and lepton fraction gradients have an important dynamical
role. This stands in contrast to the case of thin accretion discs
where only vertical buoyancy is dynamically relevant. The impact of
buoyancy on the linear growth of the MRI has been studied by
\citet{balbus94, menou04, masada06, masada07}. They showed that the
diffusion of heat and leptons driven by neutrinos alleviates the
otherwise stabilising effect of a stable stratification, such that the
MRI can still grow sufficiently fast. Neutrino radiation also induces
a high viscosity, which can limit the growth of the MRI if the initial
magnetic field is too low \citep{guilet15}.

How the non-linear phase of the MRI is affected by the radial buoyancy
and neutrino radiation is much less known. The impact of buoyancy on
the MRI has been studied in numerical simulations of "semi-global"
models \citep{obergaulinger09, masada15}, but these simulations did not
take into account the impact of neutrinos as a source of thermal
diffusion and viscosity \citep[although it has an important impact,
see][]{guilet15}. The pioneering simulations of
\citet{obergaulinger09} considered a small part of a PNS but were
not purely local in that they included global gradients, in particular
of entropy and density. They investigated the impact of buoyancy
driven by a radial entropy gradient, but artefacts induced by the
radial boundary conditions prevented an adequate study of the impact
of buoyancy on the non-linear phase of the MRI. Indeed, soon after the
MRI reached non-linear amplitudes they observed a flattening of the
entropy gradient with a strong gradient only present in a thin layer
near the radial boundaries. The radially global (but vertically local)
model of \cite{masada15} may not suffer from such
boundary artefacts (and could in principle describe meaningfully the
evolution of the global radial profiles), but the radial and azimuthal
resolution that they could afford was in our opinion much too coarse for an accurate
study of the MRI.

In order to avoid these shortcomings and to reduce the computational
cost (thus allowing an exploration of parameter space with 3D
simulations), we used a more simplified setup. We performed shearing
box simulations in the Boussinesq approximation. This approximation
takes into account the buoyancy effects due to entropy/lepton fraction
gradients, but neglects other compressibility effects. As will be
argued a posteriori, this assumption is very well justified since
magnetic and kinetic energies remain small compared to the thermal
energy. Our approach retains the effects of global gradients of
angular frequency, entropy and lepton fraction through shear and
buoyancy, but is purely local in the sense that the volume averages of
these quantities are kept fixed and that the density gradient is
neglected. It therefore retains less global effects than the
"semi-global" simulations by \citet{obergaulinger09}. This seems
necessary, however, to have clean shear-periodic boundary conditions
which avoid numerical artefacts at the grid boundary. Choosing a box
size smaller than the density scale-height ensures that this local
approach is approximately justified.

The paper is organized as follows. In Section~\ref{sec:setup}, we
describe the physical and numerical setup considered. The results are
then presented in Section~\ref{sec:linear} for the linear exponential
growth phase of the MRI, and in Section~\ref{sec:nonlinear} for the
saturated non-linear phase that follows. Finally we discuss the
validity of our assumptions in Section~\ref{sec:discussion} and draw
conclusions in Section~\ref{sec:conclusion}. A linear analysis is also
presented in Appendix~\ref{sec:linear_analysis} and
\ref{sec:hydro_linear_analysis}, the results of which have been used
in Sections~\ref{sec:linear} and \ref{sec:nonlinear}. Finally, a
convergence test of the simulations is described in
Appendix~\ref{sec:convergence}.

\section{Numerical setup}
\label{sec:setup}

\subsection{Governing equations}
The simulations performed in this article are designed to represent a
small part of a fast rotating PNS located in the equatorial plane at a
radius of around $20\,{\rm km}$. We describe the local dynamics in the
framework of a Cartesian shearing box \citep[e.g.][]{goldreich65}. The
coordinates $x$, $y$, and $z$ represent the radial, azimuthal and
vertical direction, respectively, and the corresponding unit vectors
are $\bm{e_x}$, $\bm{e_y}$, and $\bm{e_z}$. The angular frequency
vector points in $z$ direction, \ie $\bm{\Omega}=\Omega\, \bm{e_z}$,
while gravity is in $-x$ direction, \ie $\bm{g}=-g\,\bm{e_x}$. At the
location of the box considered here, the neutrinos are in the
diffusive regime \citep{guilet15}, and their effects on the dynamics
can appropriately be described by a viscosity $\nu$ as well as thermal
and lepton number diffusivities. For the sake of simplicity, we assume
in this work that the thermal and lepton number diffusivities are
equal and denoted by $\chi$, which allows us to describe the buoyancy
associated with both entropy and lepton number gradients with the use of
a single buoyancy variable (see below). For the sake of conciseness,
we refer to $\chi$ as the thermal diffusivity. We also considered the
effects of a magnetic diffusivity $\eta$ in our study.

As discussed in the introduction, buoyancy is described in the
framework of the Boussinesq approximation, \ie the buoyancy force is
the only compressibility effect taken into account. The Boussinesq
approximation holds if both the flow and Alfv\'en velocity are much
less than the sound speed (incompressible flow), and if the density
perturbations associated with the buoyancy are small. The
approximation further assumes a uniform background density $\rho_0$,
\ie one neglects the radial density gradient. The validity of these
assumptions will be assessed in Section~\ref{sec:boussinesq_validity}.

In the shearing box approximation, the diffusive MHD Boussinesq
equations read in a frame rotating with the centre of the box with an
angular frequency $\Omega$
\begin{eqnarray}
  \label{eq:base1}
  \partial_t \bm{v}+\bm{v}\cdot \bm{\nabla} \bm{v} 
   &=& - \frac{1}{\rho_0}\bm{\nabla}\Pi
       + \frac{1}{4\pi\rho_0}(\bm{\nabla}\times \bm{B})\times\bm{B},\\
  \nonumber
   & & - 2\bm{\Omega}\times\bm{v} + 2q \Omega^2 x\bm{e_x}
       - N^2\theta \bm{e_x} + \nu\bm{\Delta}\bm{v}\\
  \label{eq:base2}
  \partial_t \bm{B} &=& \bm{\nabla}\times(\bm{v}\times\bm{B})
                        + \eta\bm{\Delta{B}},\\
  \label{eq:base3}
  \partial_t \theta + \bm{v}\cdot \bm{\nabla} \theta 
   &=& v_x + \chi \Delta \theta, \\
  \label{eq:base4}  \nabla \cdot \bm{v}&=&0,\\
  \label{eq:base5}  \nabla \cdot \bm{B}&=&0,
\end{eqnarray}
where $\bm{v}$ and $\bm{B}$ are the flow velocity and the magnetic
field, respectively. The buoyancy variable
\begin{equation}
  \theta \equiv -\frac{g}{N^2} \frac{\delta \rho}{\rho_0}
  \label{eq:theta}
\end{equation} 
describes the density perturbation $\delta \rho$ due to the joint effect of entropy and lepton number perturbations,
and has the dimension of a length. The gradient of the pressure perturbation
  $\bm{\nabla}\Pi$ is obtained from the constraint of a divergence-free
flow field (equation~\ref{eq:base4}). The quantity
\begin{equation}
  q \equiv - \dd \log \Omega / \dd \log r
  \label{eq:q}
\end{equation}
measures the local shear (we assume $q=1.25$ in all simulations). The
square of the Brunt-V\"ais\"al\"a frequency $N$ is defined by
\begin{equation}
  N^2 \equiv - \frac{g}{\rho}\left\lbrack
               \frac{\partial \rho}{\partial S}\Big|_{P,Y_e}
               \frac{\dd S}{\dd r} +
               \frac{\partial \rho}{\partial Y_e}\Big|_{P,S}
               \frac{\dd Y_e}{\dd r}  \right\rbrack,
  \label{eq:BV}
\end{equation}
where $\rho$, $S$, $Y_e$, and $P$ are the density, entropy,
  electron fraction, and pressure, respectively. A stationary
solution of these equations is:
\begin{eqnarray}
  \bm{v_0} &=& - q\Omega x \bm{e_y}, \\
  \bm{B_0} &=& B_0 \bm{e_z}, \\
  \theta_0 &=& 0.
\end{eqnarray}
We initialised the numerical simulations with this stationary solution,
to which we imposed random velocity and magnetic field perturbations
(as described in more detail in next subsection).

\subsection{Physical parameters and dimensionless numbers}
We chose the physical parameters of the simulations to represent the
fast rotating PNS model studied by \citet{guilet15} at a radius of
$\simeq 20\,{\rm km}$ with a density
$\rho_0 = 2\times10^{13}\, {\rm g\,cm^{-3}}$, a viscosity
$\nu = 10^{10}\,{\rm cm^2\,s^{-1}}$, and a rotation angular frequency
$\Omega = 10^3\,{\rm s^{-1}}$. The box size was set to
$(L_x,L_y,L_z) = (4,4,1)\,{\rm km}$. The radial width of the box is
somewhat smaller than the density scale height $\sim 8\,{\rm km}$, \ie
the neglect of the background density gradient is roughly justified
(see Sections~\ref{sec:boussinesq_validity} and \ref{sec:box_size} for
a discussion). The aspect ratio of the box, elongated in the
horizontal directions compared to the vertical direction, was chosen
to ensure that the box fits the most unstable parasitic modes growing
on MRI channel modes \citep{goodman94, pessah09, latter09,
  obergaulinger09, pessah10}.

Given these values for the box size and for the viscosity, the
dimensionless {\it Reynolds number} (characterizing the importance of
viscosity over the largest vertical scale in the box) is
\begin{equation}
  R_e \equiv \frac{L_z^2\Omega}{\nu} = 10^{3}.
  \label{eq:Re}
\end{equation}

According to \citet{masada07}, the thermal diffusivity $\chi$ is
larger than the neutrino viscosity $\nu$ by a factor $10-100$.  To
investigate its impact on the results we chose two different values
for the thermal diffusivity, namely a "high" one
$\chi = 10^{11}\,{\rm cm^2\,s^{-1}}$, and a ten times lower one
$\chi=10^{10}\,{\rm cm^2\,s^{-1}}$ referred to as "low diffusion" in
the following. These values of the thermal diffusivity correspond to a
dimensionless {\it Prandtl number} (characterizing the relative
importance of viscosity and thermal diffusion)
\begin{equation}
  Pr \equiv \nu/\chi = 0.1 \,\, \mathrm{and} \,\, 1,
  \label{eq:Pr}
\end{equation}
respectively. Hence, the dimensionless {\it Peclet number}
(characterizing the relative importance of advection and thermal
diffusion) has a value of
\begin{equation}
  P_e \equiv \frac{L_z^2\Omega}{\chi} = 100 \,\, \mathrm{and} \,\, 1000
  \label{eq:Pe}
\end{equation}
for the high and low diffusion case, respectively.

The resistivity $\eta$ is expected to be many orders of magnitude
smaller than the viscosity $\nu$, but numerical constraints do not
allow one to perform numerical simulations for such values of the
resistivity.  In our study, we used a resistivity
$\eta = 2.5\times 10^{9}\,{\rm cm^2\,s^{-1}}$, which corresponds to a
{\it magnetic Prandtl number} characterizing the relative importance
of viscosity and resistivity of
\begin{equation}
  P_m \equiv \nu/\eta = 4. 
  \label{eq:Pm}
\end{equation}
This much too small value is a compromise between the computing
time constraints and the wish to simulate flows in the high magnetic Prandtl regime appropriate to a PNS
\citep[$P_m \sim 10^{13}$ ;][]{thompson93, masada07}. Note that this regime is unusual since the magnetic Prandtl number is ordinarily very small for stars, most parts of accretion discs and liquid laboratory metals. The high value of the magnetic Prandtl number in a PNS is due to the neutrinos inducing a very large viscosity while the resistivity remains very small. The dependence of the results on the magnetic Prandtl number will be investigated in a forthcoming publication. The
corresponding {\it magnetic Reynolds number}, which characterizes the
relative importance of magnetic advection to magnetic diffusion
(resistivity), is
\begin{equation}
  R_m \equiv \frac{L_z^2\Omega}{\eta} = 4\times10^{3}
  \label{eq:Rm}
\end{equation}
in our study.

We considered a single value of the magnetic field strength, whose
value $B_0=1.6\times10^{13}\,{\rm G}$ lies within the viscous regime
described by \citet{guilet15} (see left panel of their Fig.~10). A
dimensionless number characterizing the strength of the magnetic field
is defined as
\begin{equation}
  \beta \equiv \frac{L_z^2\Omega^2}{v_A^2} = 10^4,
  \label{eq:def_beta}
\end{equation}
where $v_A = B/\sqrt{4\pi\rho_0}$ is the Alfv\'en speed associated with
the background vertical magnetic field. The impact of viscosity on the
linear growth of the MRI is determined by the {\it viscous Elsasser
  number}
\begin{equation}
  E_\nu \equiv \frac{v_{A}^2}{\nu\Omega} = \frac{R_e}{\beta} = 0.1.
  \label{eq:Enu}
\end{equation}
Since $E_\nu<1$, our simulations belong to the viscous regime, where
the growth rate of the MRI is significantly reduced by neutrino viscosity
compared to ideal MRI \citep[e.g.][and references
therein]{guilet15}. Similarly, the impact of resistivity on the
linear growth of the MRI is quantified by the {\it resistive Elsasser
  number}
\begin{equation}
  E_\eta \equiv \frac{v_{A}^2}{\eta\Omega} = \frac{R_m}{\beta} = 0.4.
  \label{eq:Eeta}
\end{equation}
Contrary to what would be expected under realistic conditions, the
impact of resistivity on the linear MRI is therefore not negligible in
our setup (since $E_\eta<1$). Finally, the impact of thermal diffusion
on the linear growth of the MRI is quantified by the {\it thermal
  Elsasser number}
\begin{equation}
  E_\chi \equiv \frac{v_{A}^2}{\chi\Omega}
  = \frac{P_e}{\beta} = 0.01 \,\, \mathrm{and} \,\, 0.1,
  \label{eq:Echi}
\end{equation}
for the high and low diffusion case, respectively. This shows
that thermal diffusion is expected to play an important role in the
MRI linear growth, as will be detailed in Section~\ref{sec:linear}.

Furthermore, we varied the Brunt-V\"ais\"al\"a frequency (see
equation~\ref{eq:BV}), which characterizes the impact of buoyancy,
  performing a set of simulations with
  $N^2/\Omega^2 \in \{ -1.5, -1, -0.5, 0, 1, 2, 3, 4, 6, 8, 10 \}$
  both for the low and high thermal diffusion case.

In order to trigger the growth of the MRI, we added velocity and
magnetic field perturbations to the stationary state in the following
way. We imposed Fourier modes with a random phase and an amplitude that
follows a Kolmogorov spectrum. The spectrum is normalized such that
the kinetic and magnetic energy of the perturbations amount to $1\%$
of the energy in the background magnetic field. The results in the non-linear phase are not
very sensitive to the initial conditions, which influence the duration
of the initial exponential growth phase (higher amplitude
perturbations lead to a shorter growth phase).

Note that the simulations presented in this paper, meant to describe
the flow in a fast rotating neutron star at a radius of
$20\,{\rm km}$, can be rescaled to other physical situations sharing
the same values of the dimensionless numbers. For example, the
moderately rotating PNS model studied by \citet{guilet15} shares the
same dimensionless numbers at a radius of $10\,{\rm km}$ with the
following parameters: $\Omega=200\,{\rm s^{-1}}$,
$\nu = 2\times10^9\,{\rm cm^2\,s^{-1}}$,
$\rho = 10^{14}\,{\rm g\,cm^{-3}}$, $B_0 = 7\times 10^{12}\,{\rm G}$,
box size $(L_x,L_y,L_z) = (4,4,1)\,{\rm km}$, Prandtl numbers
${\rm Pr} = 0.1-1$, and magnetic Prandtl number ${\rm P_m} = 4$. The
Brunt-V\"ais\"al\"a frequencies considered then range from
$N^2 = -6\times10^{4}\,{\rm s^{-2}}$ to $4\times10^{5}\,{\rm s^{-2}}$
in that situation.

\subsection{Units}
Throughout the paper, we normalise our results using the vertical size
of the domain $L_z=1\,{\rm km}$, the angular frequency
$\Omega = 10^3\,{\rm s^{-1}}$, and the density
$\rho_0 = 2\times 10^{13}\,{\rm g\,cm^{-3}}$. Time is therefore
measured in units of $1\,{\rm ms}$, velocity in units of
$10^8 \,{\rm cm\,s^{-1}}$, the magnetic field in units of
$1.6\times10^{15}\,{\rm G}$, the energy density in units of
$2\times 10^{29}\,{\rm erg\,cm^{-3}}$, the specific energy in units of
$10^{16}\,{\rm erg/g}$, the energy injection rate density in units of
$2\times10^{32}\,{\rm erg\,s^{-1}\,cm^{-3}}$, and the specific energy
injection rate in units of $10^{19}\,{\rm erg\,s^{-1}\,g^{-1}}$.

\subsection{Numerical methods}
In order to solve the Boussinesq MHD
equations~(\ref{eq:base1})--(\ref{eq:base5}), we used the
pseudo-spectral code \textsc{snoopy} \citep{lesur05,lesur07}.
\textsc{snoopy} solves the 3D shearing box equations using a spectral
Fourier method, where the shear is handled through the use of a
shearing wave decomposition (with time varying radial wavevector) and
a periodic remap procedure. Nonlinear terms are computed with a
pseudo-spectral method using the 2/3 dealiasing rule. The time
integration is performed using an implicit procedure for the diffusive
terms, while other terms use an explicit 3rd order Runge Kutta
scheme. \textsc{snoopy}
is parallelized using both MPI and OpenMP techniques. \textsc{snoopy}
has been used in the past to study the MRI (in the absence of
buoyancy) by \citet{lesur07}, \citet{longaretti10}, \citet{rempel10},
and \citet{lesur11}. The Boussinesq implementation has been used to
study vertical convection and the subcritical baroclinic instability
in an accretion disc by \citet{lesur10a} and \citet{lesur10b}.

Most simulations presented in this paper were performed using a
standard grid resolution of $(n_x,n_y,n_z)=(256,128,64)$ zones. A
convergence study, presented in Appendix~\ref{sec:convergence}, shows
that this resolution is sufficient to resolve the dissipative scales
and that the results are consistent with those obtained at a twice
higher resolution. All our simulations with $N^2/\Omega^2\geqslant
-0.5$ were performed with the standard resolution. At more negative
values of $N^2$, however, we observed that the more vigorous
turbulence led to smaller dissipative scales. Therefore, we increased
the resolution to $(n_x,n_y,n_z) = (384,192,96)$ for $N^2/\Omega^2 =
-1$ (both for $P_e=100$ and $P_e=1000$), and to $(n_x,n_y,n_z) =
(512,256,128)$ for the simulation with $N^2/\Omega^2 = -1.5$ and
$P_e=100$. We note that in all our simulations the resolution is twice
lower in the azimuthal direction than in the radial or vertical
direction. This is rather usual in numerical studies of the MRI,
because the presence of shear makes structures more elongated in the
azimuthal direction. We have checked that our results are not affected
by the lower azimuthal resolution.

\subsection{Turbulent energies and energy injection rates}
We define the kinetic, magnetic, and thermal energy densities as
\begin{equation}
  E_{\rm kin} \equiv \frac{1}{2}\rho_0\bm{u}^2,
  \label{eq:def_Ekin}
\end{equation}
\begin{equation}
  E_{\rm mag} \equiv \frac{1}{8\pi}B^2,
  \label{eq:def_Emag}
\end{equation}
and
\begin{equation}
  E_{\rm th} \equiv \frac{1}{2}\rho_0N^2\theta^2,
  \label{eq:def_Eth}
\end{equation}
respectively, where the velocity $\bm{u}$ is defined with respect
  to the equilibrium shear profile $\bm{v_0}$ according to
  $\bm{u} \equiv \bm{v} - \bm{v_0}$.  For brevity, we will refer to $\bm{u}$ as the velocity, and to energy densities as energies.
  
 We further denote the average of a quantity $A$ over the whole computational volume by
$\langle A \rangle$. The sum of the volume averaged kinetic and magnetic energies obeys the following conservation equation
\begin{equation}
  \frac{\dd \langle E_{\rm kin}\rangle+\langle E_{\rm mag}\rangle}{\dd t} =
  q\Omega\left\lbrack \langle R_{xy}\rangle + \langle M_{xy}\rangle
  \right\rbrack + \langle W\rangle - \langle \epsilon_{\nu}\rangle
  - \langle \epsilon_\eta\rangle,
  \label{eq:econs}
\end{equation}
where the Reynolds stress
\begin{equation}
  R_{xy} \equiv \rho_0 u_xu_y
  \label{eq:def_Reynolds}
\end{equation}
and the Maxwell stress
\begin{equation}
  M_{xy} \equiv \frac{-B_xB_y}{4\pi}
  \label{eq:def_Maxwell}
\end{equation}
are the angular momentum fluxes in the radial direction due to the
flow and the magnetic field, respectively. The sum of
  $q\Omega \langle R_{xy} \rangle$ and
  $q\Omega \langle M_{xy} \rangle$ gives the energy density injection
  rate into turbulent flow extracted from the shear energy of
  the flow by Reynolds and Maxwell stresses.
\begin{equation}
  W \equiv -N^2 \rho_0 \theta u_x
  \label{eq:def_buoyancy_work}
\end{equation}
is the work done by the buoyancy force per unit of time and per volume,
  \ie $\langle W\rangle$ in equation~\ref{eq:econs} represents the energy
  exchange rate from thermal to kinetic energy through
  radial transport of entropy and lepton number caused by buoyancy.
  The remaining quantities in equation~\ref{eq:econs},
  $\epsilon_\nu\equiv \rho_0 \nu |\bm{\nabla}\times \bm{u}|^2$ and
  $\epsilon_\eta\equiv \eta |\bm{\nabla}\times \bm{B}|^2/4\pi$, are
  the energy dissipation rates due to the viscosity and the resistivity, respectively.

In a time average sense, the total energy injection rate
into turbulence, $ q\Omega\left\lbrack \langle R_{xy}\rangle + \langle
  M_{xy}\rangle \right\rbrack + \langle W\rangle$, is equal to the sum
of the viscous and resistive energy dissipation rates
$\langle\epsilon_\nu + \epsilon_\eta\rangle$.

\section{Linear growth phase}
\label{sec:linear}

\begin{figure*}
  \centering
  \includegraphics[width=2\columnwidth]{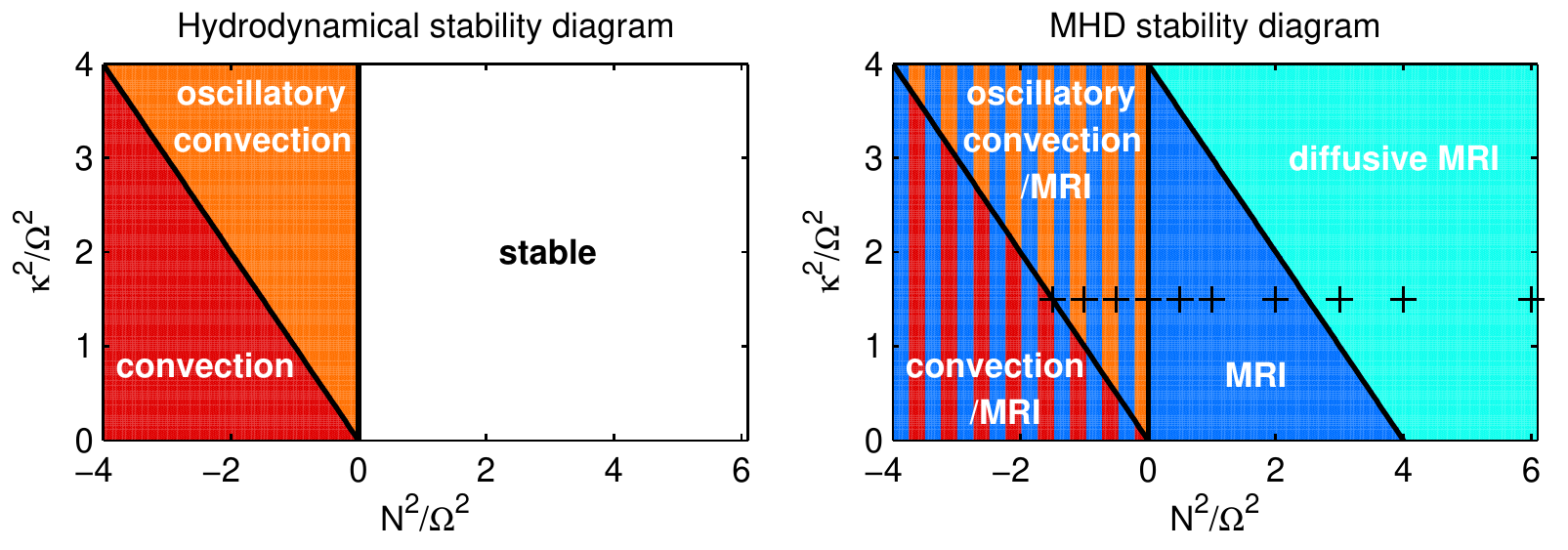}
  \caption{Stability diagram as a function of the square of the
    Brunt-V\"ais\"al\"a frequency $N^2$ (characterizing the influence
    of buoyancy) and the square of the epicyclic frequency $\kappa^2$
    (characterizing the effect of rotation and shear). The left panel
    applies to hydrodynamic flows and shows the regions unstable
    against classical convection (red) and oscillatory convection
    enabled by thermal diffusion (orange). The right panel applies to
    MHD flows and shows the region unstable against the MRI (blue)
    including the also convectively unstable regions (red and orange
    regions with blue stripes) and the diffusive MRI enabled by thermal
    diffusion (turquoise). The plus signs give the parameters of our
    simulations (those for $N^2/\Omega^2 = 8$ and 10 are not shown).}
  \label{fig:stability_diagram}
\end{figure*}

\begin{figure*}
  \centering
  \includegraphics[width=\columnwidth]{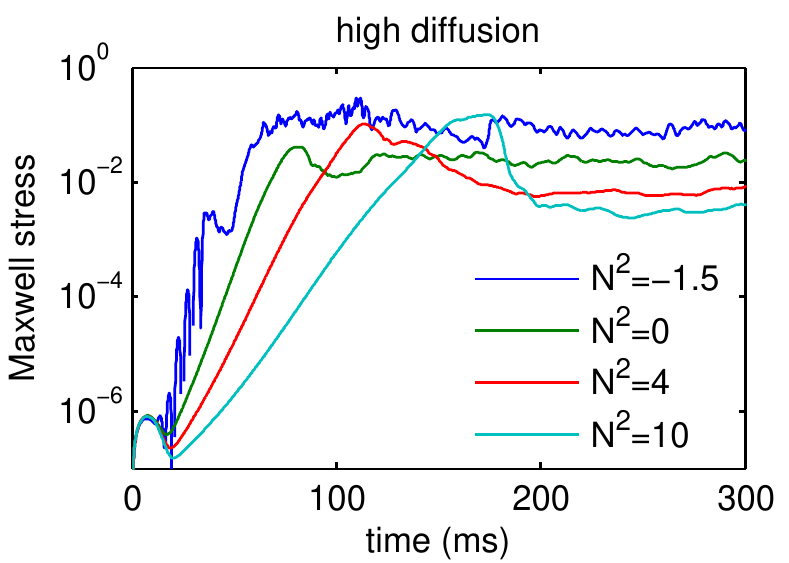}
  \includegraphics[width=\columnwidth]{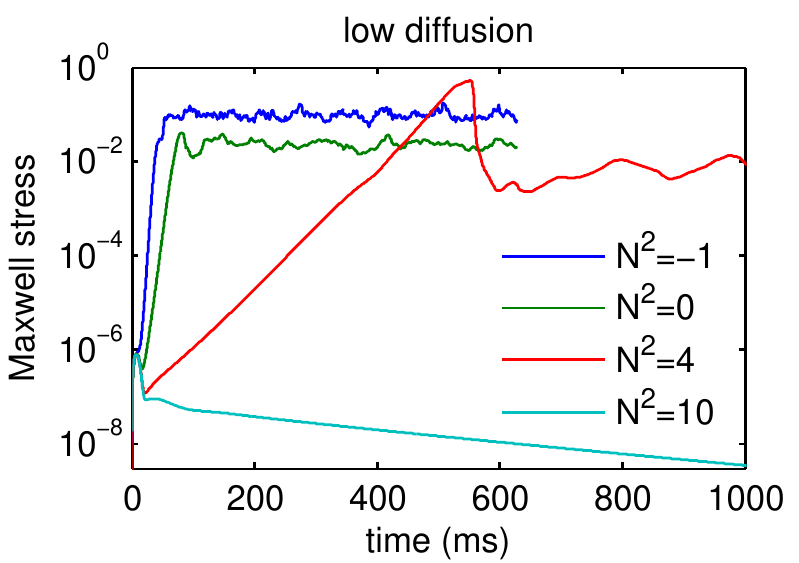}
  \caption{Time evolution of the box averaged Maxwell stress for
    simulations with different stratifications, \ie different values of
    $N^2/\Omega^2$, for high ($P_e=100$; left panel) and low ($P_e=
    1000$; right panel) thermal diffusion. Note that the time interval
    is different in the two panels.}
  \label{fig:maxwell_vs_time}
\end{figure*}

\begin{figure*}
  \centering
  \includegraphics[width=\columnwidth]{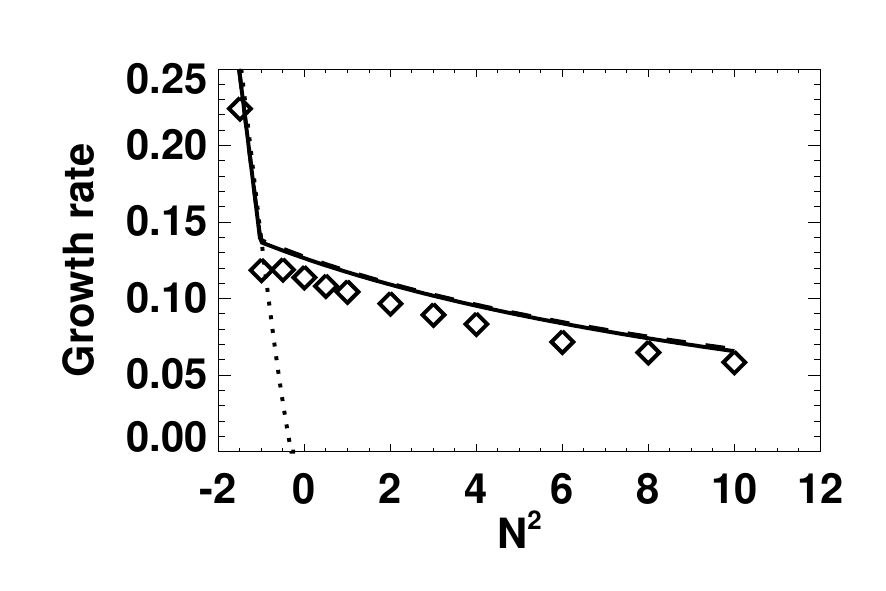}
  \includegraphics[width=\columnwidth]{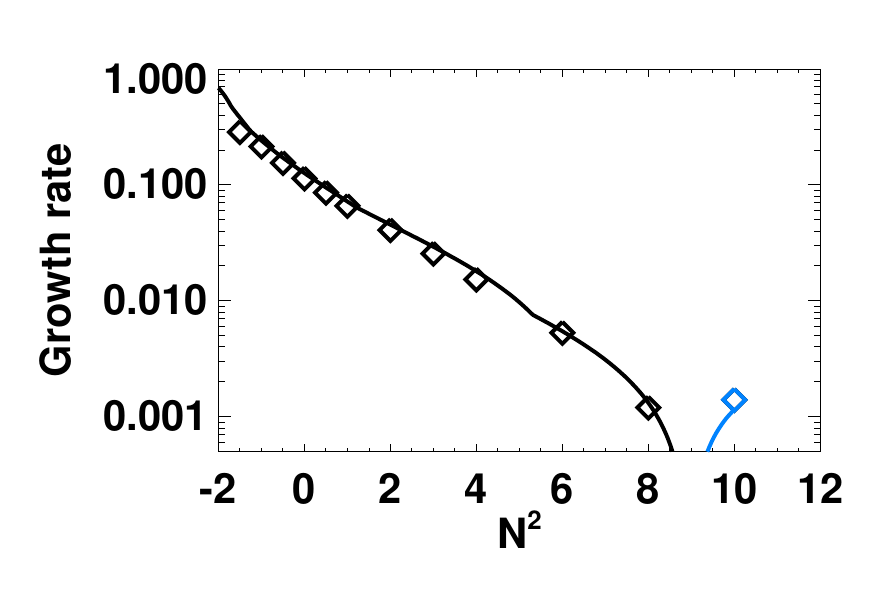}
  \caption{Comparison of the growth rates as a function of the strength
    of buoyancy as measured by $N^2/\Omega^2$ obtained from our
    numerical simulations (diamonds) and our linear analysis (solid
    lines; see Appendix~\ref{sec:linear_analysis} and
    \ref{sec:hydro_linear_analysis}) for high ($P_e=100$; left panel)
    and low ($P_e= 1000$, right panel) thermal diffusion. The dotted
    line gives the prediction of the linear growth rate in a
      hydrodynamic flow (Appendix~\ref{sec:hydro_linear_analysis}).
    Blue colour is used when the growth rate is negative, \ie when it
    gives the damping rate. We normalized both the growth rate $\sigma$
    and the Brunt-V\"ais\"ala frequency $N$ by the angular frequency
    $\Omega$.}
  \label{fig:growth_rate}
\end{figure*}

Let us start this section by discussing stability conditions in the
presence of buoyancy and differential rotation both in hydrodynamic
and MHD flows. We restrict this discussion to the special case considered in this paper: a local box in the equatorial plane of the PNS, where the angular frequency, entropy and composition depend only on radius. We further restrict the discussion to axisymmetric perturbations\footnote{Due to the presence of shear, non-axisymmetric perturbations cannot lead to exponential growth but can nonetheless be transiently growing. While this may not be considered as leading to genuine linear instability, it can be important for the non-linear turbulent dynamics.} and to flows which are stable
against the hydrodynamic shear instability, \ie complying with the
Rayleigh stability criterion $\kappa^2 > 0$, where $\kappa$ is the
epicyclic frequency defined by
\begin{equation}
  \kappa^2 \equiv \frac{1}{r^3}\frac{\dd (r^4\Omega^2)}{\dd r}.
  \label{eq:def_kappa}
\end{equation}

We first consider stability conditions in the absence of any diffusive
process. If
\begin{equation}
  \kappa^2 + N^2 < 0
  \label{eq:hydro_stability}
\end{equation}
the flow is convectively unstable  because the Solberg-H\o iland stability criteria is violated (red region in
Fig.~\ref{fig:stability_diagram}). In the presence of magnetic field
the condition reads
\begin{equation}
  \kappa^2 + N^2 < 4\Omega^2.
  \label{eq:MHD_stability}
\end{equation}
This criterion is less restrictive than the hydrodynamic one due to
the appearance of the MRI for $0 < \kappa^2 + N^2 < 4\Omega^2$ (dark
blue region, and red and orange regions with blue stripes in the right
panel of Fig.~\ref{fig:stability_diagram}).  The classical MRI (\ie
in the absence of buoyancy effect) is obtained for $N^2 = 0$ and $0<
\kappa^2 < 4\Omega^2$.

When thermal diffusion is present (but no resistivity and viscosity),
the instability region is more extended for two reasons. First,
thermal diffusion alleviates the stabilising effect of buoyancy on the
MRI \citep{acheson78, menou04, masada07}. As a consequence the MRI
instability criterion becomes $0 < \kappa^2 < 4\Omega^2$, \ie the MRI
unstable region now comprises a region that is stable according to
equation~(\ref{eq:MHD_stability}) (turquoise region in the right panel of
Fig.~\ref{fig:stability_diagram}).

Another consequence of thermal diffusion is the appearance of a new
hydrodynamic instability \citep{klahr14, lyra14}, which we refer to as
oscillatory convection. It occurs for $N^2<0$ in a parameter regime
that is stable according to equation~(\ref{eq:hydro_stability}) (orange
region in Fig.~\ref{fig:stability_diagram}). Oscillatory convection
can be considered to be an analog of semi-convection (in which thermal
diffusion causes the flow to be unstable in the presence of an
otherwise stabilising composition gradient), rotation assuming the
role of the stabilising composition gradient in the case of
semi-convection.

The parameter regime explored with our simulations is shown with plus
signs in Fig.~\ref{fig:stability_diagram}. We varied the
Brunt-V\"ais\"al\"a frequency in the range $-1.5 <N^2/\Omega^2 < 10$
and fixed the epicyclic frequency to a value $\kappa^2/\Omega^2 =
1.5$.  Hence, we explored three regimes: non-diffusive MRI for
$0<N^2<2.5$, diffusive MRI (\ie allowed by thermal diffusion) for
$N^2> 2.5$, and a mixed regime where MRI and oscillatory convection
can co-exist for $-1.5<N^2<0$. Note that in none of our simulations
the flow is unstable to {\bf axisymmetric} classical (non-oscillatory) convection.

Figure~\ref{fig:maxwell_vs_time} shows the time evolution of the
Maxwell stress averaged over the computational box. After a linear
phase of exponential growth, the Maxwell stress saturates in the
non-linear phase. In this section we analyse only the linear phase and
postpone the analysis of the non-linear phase to next section. All but
two simulations show a non-oscillatory growth, which is due to the MRI
modified by buoyancy. For low thermal diffusion ($P_e = 1000$; right
panel), the growth rate varies by orders of magnitudes and becomes
extremely small for large positive values of $N^2$, up to the point
that no growth occurs at all for $N^2/\Omega^2=10$. By contrast, for
high thermal diffusion ($P_e = 100$; left panel), the effect of
buoyancy on the growth of the MRI is much less pronounced.  The MRI
can still grow in a short time even for large values of $N^2$. A good
agreement is obtained between the growth rates obtained from our
numerical simulations and the predictions from the linear analysis
presented in Appendix~\ref{sec:linear_analysis}
(Fig.~\ref{fig:growth_rate}).

Our results are in qualitative agreement with those of
\citet{masada07}, who showed that the MRI can grow unimpeded by stable
buoyancy if the thermal diffusivity is sufficiently large compared to
the viscosity, \ie if $\chi > \nu N/\Omega$. This condition is indeed
verified for high thermal diffusion ($P_e=100$ and $\chi/\nu=10$), in
which case the growth of the MRI is only mildly affected by
buoyancy. By contrast, the growth of the MRI is drastically suppressed
in the case of low thermal diffusion and large $N^2/\Omega^2$
($P_e=1000$ and $\chi=\nu$), \ie if the above condition is not
fulfilled. Note that, even for a high diffusivity, the MRI growth
rates are significantly smaller than those of the ideal MRI
($\sigma/\Omega =q/2 =0.625$). This stands in contrast to the
situation considered by \citet{masada07} and is due to the fact that
we considered a magnetic field strength that lies in the viscous
regime, where the MRI growth is significantly slowed down by neutrino
viscosity \citep{guilet15}.

One of our simulations with unstable buoyancy and high thermal
diffusion ($N^2/\Omega^2=-1.5$ and $P_e=100$) showed an oscillatory
exponential growth due to the oscillatory convection discussed
above. The transition of the fastest growing mode from MRI to
oscillatory convection is clearly seen in Fig.~\ref{fig:growth_rate}
for high thermal diffusion, where it happens at $N^2\ \sim -\Omega^2$
in agreement with the linear prediction. In the regime of oscillatory
convection, the growth rate predicted by the hydrodynamic linear
analysis presented in Appendix~\ref{sec:hydro_linear_analysis} matches
that of the MHD linear analysis and the simulations, clearly
demonstrating the hydrodynamic nature of the instability.

\section{Non-linear phase}
\label{sec:nonlinear}

When the exponentially growing axisymmetric (\ie y-independent)
MRI channel modes reach non-linear amplitudes, non-axisymmetric secondary (or
"parasitic") instabilities start to grow upon the primary modes
\citep{goodman94, pessah09, latter09, obergaulinger09, pessah10}. This
leads to the disruption of the channel modes, thereby ending the
exponential growth phase. From that point on the dynamics consists, in
most cases, of 3D statistically steady MHD turbulence. A somewhat
different dynamics takes place, however, in the case of low thermal
diffusion ($P_e=1000$) and a sufficiently stabilising buoyancy
$N^2/\Omega^2 \gtrsim 3$. In that case, we observed the recurrent
emergence of exponentially growing channel flows and their subsequent
disruption by parasitic instabilities. The resulting dynamics is
fundamentally non-steady due to this quasi-periodic emergence of the
large scale channel flows. For high thermal diffusion we did not
observe recurrent channel flows, but nonetheless noticed the
development of long-lived coherent flow structures in the regime where
the flow is stable against buoyancy. Therefore, we split the following
discussion of the properties of the non-linear phase into two
subsections. The first one we devote to the coherent flows (either
recurrent or statistically steady), while we discuss in the second
subsection the time and volume averaged properties of the flow during
the non-linear phase.

\subsection{Coherent flows}
\label{sec:coherent_flows}

\begin{figure}
  \centering
  \includegraphics[width=\columnwidth]{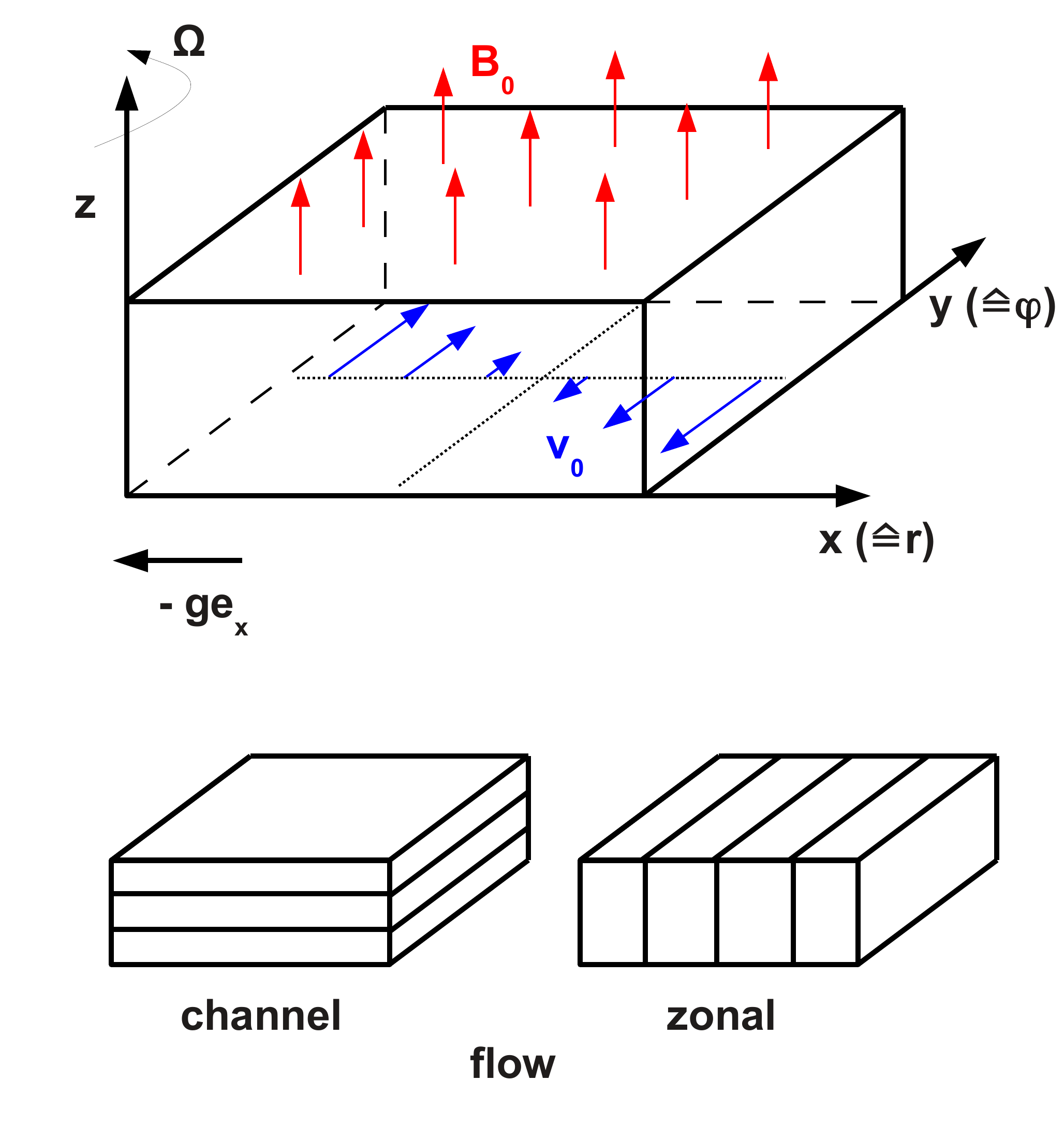}
  \caption{Schematic representation of the shearing box geometry (upper panel) and of the geometry of the coherent flows (lower panels): channel flows (left-hand lower panel) and zonal flows (right-hand lower panel).}
  \label{fig:shearing_box}
\end{figure}

\begin{figure*}
  \centering
  \includegraphics[width=\columnwidth]{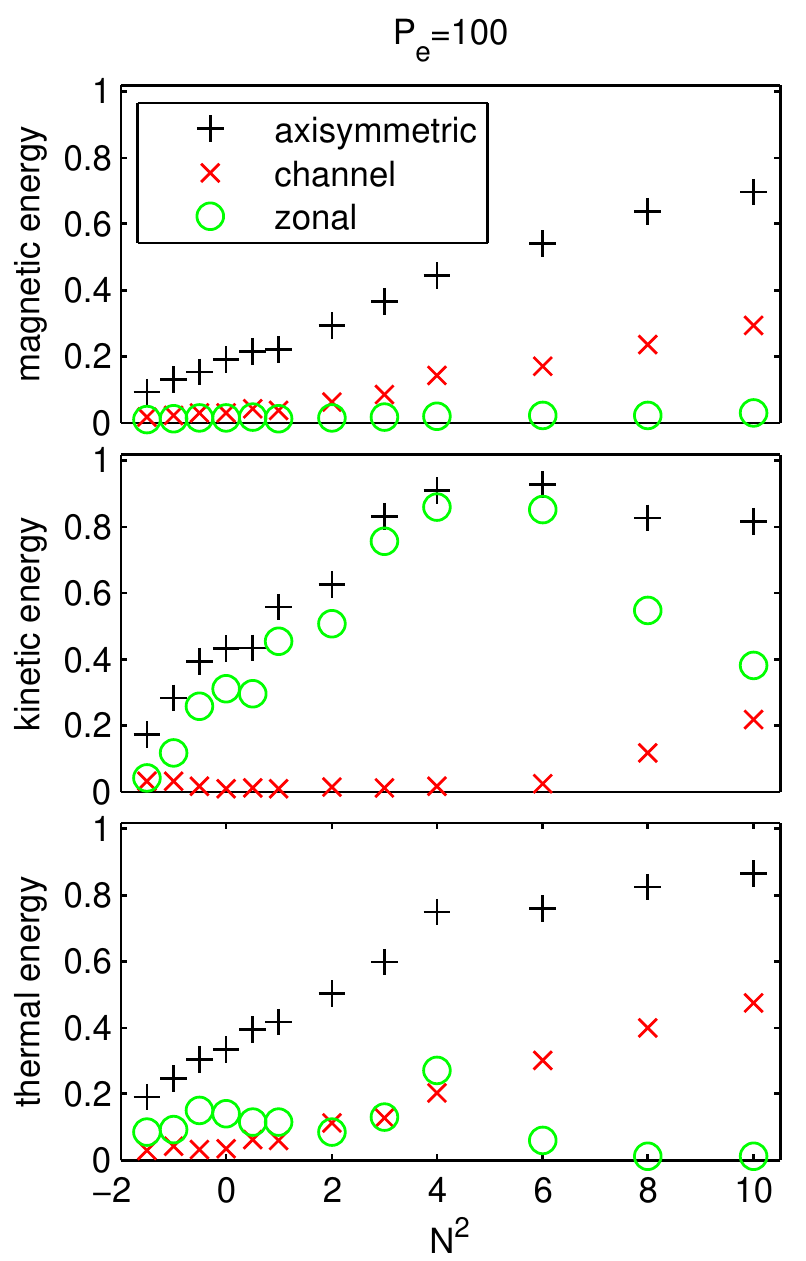}
  \includegraphics[width=0.985\columnwidth]{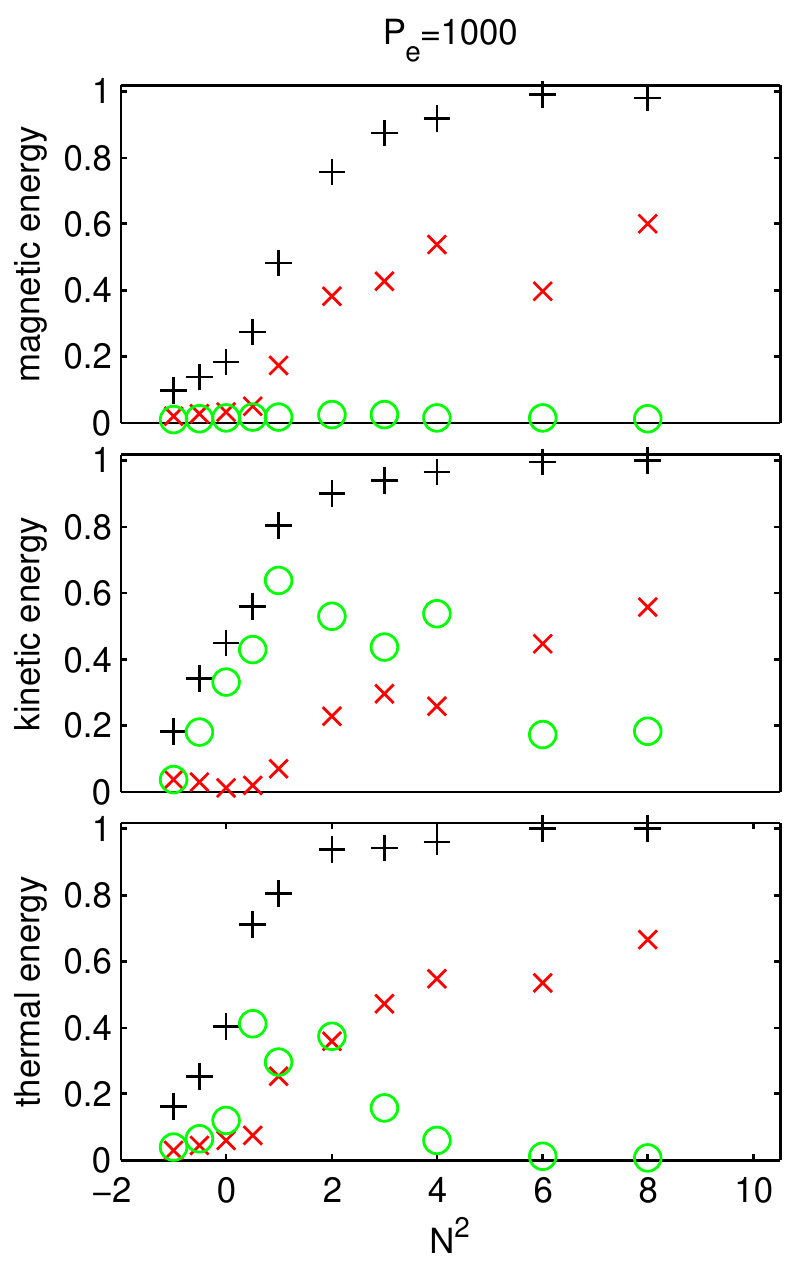}
  \caption{Fraction of magnetic (top row), kinetic
    (middle row), and thermal (bottom row) energy densities contained in
    axisymmetric structures (black $+$ symbols), channel flows (red
    $\times$ symbols), and zonal flows (green circles) for simulations
    performed with high thermal diffusion ($P_e=100$; left column) and
    low thermal diffusion ($P_e=1000$; right column).}
  \label{fig:channels_fraction}
\end{figure*}

In the stable buoyancy regime ($N^2>0$), we observe the appearance of
either long lived (for $P_e=100$) or recurrent (for $P_e=1000$)
coherent structures in the non-linear state of the MRI. Two different
types of such coherent structures can be distinguished (Figure~\ref{fig:shearing_box}): channel flows
with a vertically varying but horizontally uniform structure (\ie
  flow quantities only depend on coordinate $z$), and zonal flows
with a radially varying but azimuthally and vertically uniform
structure (\ie flow quantities only depend on coordinate $x$).

Channel flows are related to the most unstable linear modes of the MRI
in case of a vertical background magnetic field. These modes,
  which have a purely vertical wavenumber, \ie $\bm{k}=(0,0,k_z)$,
are known to be non-linear growing solutions of the MHD equations
\citep{goodman94}. They are unstable to parasitic instabilities of
Kelvin-Helmholtz or resistive tearing mode type \citep{goodman94,
  latter09, pessah09, pessah10, obergaulinger09}, the development of
which can stop the growth of the channel mode and trigger
turbulence. The typical structure of MRI channel modes consists of a
sinusoidal profile that holds for the horizontal (\ie x and y)
components of both velocity and magnetic field
\footnote{In a shearing box with shear-periodic boundary conditions,
  the horizontally averaged vertical (z) component of both velocity
  and magnetic field is conserved and uniform due to the divergence
  free condition (Eqs.~\ref{eq:base4} and \ref{eq:base5}), and has a
  value of zero and $B_0$ respectively.},
and in the presence of buoyancy for the buoyancy variable $\theta$
(equation~\ref{eq:theta}), too. As can be deduced from
Eqs.~(\ref{eq:uy_ux})--(\ref{eq:By_ux}), in MRI channel modes the
radial (x) and azimuthal (y) velocity components are typically in
phase
\footnote{In the regime where the flow is unstable to buoyancy, they
  can be also in anti-phase, \ie the associated Reynolds stress
  becomes negative (see section \ref{sec:time_average} and
  Appendix~\ref{sec:linear_analysis}).},
while the radial and azimuthal magnetic field components are typically
in anti-phase with each other
\footnote{In the regime where the flow is unstable to buoyancy, they
  can be in principle in phase too, \ie the Maxwell stress would be
  negative. This was, however, not observed for the flow parameters
  considered in this study.}.
They are shifted also by a quarter of a wavelength with respect to the
horizontal (x,y) velocity components, whereas the buoyancy variable
$\theta$ is in phase with these components (equation~\ref{eq:theta_ux}).

Zonal flows have been observed in shearing box simulations of MRI
turbulent stratified accretion discs \citep{johansen09, simon12,
  simon14}. They mostly possess a radially varying azimuthal velocity,
which is associated with a radial variation of the pressure. In the
presence of a mean vertical magnetic field, a radial variation of the
vertical magnetic field has been reported recently too \citep{bai14b,
  bai14c}.

Figure~\ref{fig:channels_fraction} illustrates the occurrence of
  the different types of coherent flows as a function of the
  Brunt-V\"ais\"al\"a frequency by showing the fractions of kinetic,
  magnetic, and thermal energy densities contained in
  axisymmetric structures, channel flows, and zonal flows,
  respectively. To obtain the energy densities contained in
  axisymmetric structures we first averaged the magnetic field,
  velocity, and buoyancy variable in azimuthal (y) direction. Then we
  used these averages to compute the respective energy densities averaging
  over the box and over time. Instead, we first averaged the flow
  variables horizontally (\ie both in azimuthal (y) and radial (x)
  direction) to compute the energy densities contained in channel flows.
  Finally, for the zonal flows, we first averaged the flow variables
  in vertical (z) and azimuthal (y) direction.

Figure~\ref{fig:channels_fraction} demonstrates that the flow becomes
more axisymmetric for increasing $N^2$, \ie for flows that are
  more stable to buoyancy. This holds for the magnetic, kinetic, and
thermal energies at both low and high thermal diffusion
but is more pronounced at low thermal diffusion, in qualitative
agreement with the idea that high thermal diffusion somewhat
alleviates the influence of a stabilising buoyancy. The prominence of
channel flows also increases monotonically with $N^2$ for all
three energies and at both thermal diffusivities. This behaviour is
again more pronounced at low thermal diffusion, where the channel
flows contain up to about $50\%$ of the energy. As already mentioned
the channel flows have a different time behaviour at $P_e=100$ and
$P_e=1000$. At high thermal diffusion they are approximately steady
(see Section~\ref{sec:steady_channels}), while at low thermal
diffusion they show recurrent phases of exponential growth and
disruption (see Section~\ref{sec:recurrent_channels}).

Zonal flows stick out most clearly regarding the kinetic energy
  (remarkably, their fraction can reach up to $90\%$), while they
amount to a lesser fraction of the thermal energy, and an almost
negligible fraction of the magnetic energy. Their dependence on $N^2$
is more complex than that of the channel flows. While their importance
first increases with $N^2$ (up to $N^2=4-6$ at $P_e=100$, and $N^2=2$
at $P_e=1000$), it starts to decrease at still higher $N^2$. As for
the channel flows, zonal flows have a different time behaviour at high
and low diffusion. They are approximately steady and long-lived at
$P_e=100$ (Section~\ref{sec:steady_zonal}), but recurrent at
$P_e=1000$ (Section~\ref{sec:recurrent_channels}).

\subsubsection{Quasi-stationary channel flows}
\label{sec:steady_channels}

\begin{figure*}
  \centering
  \includegraphics[width=\columnwidth]{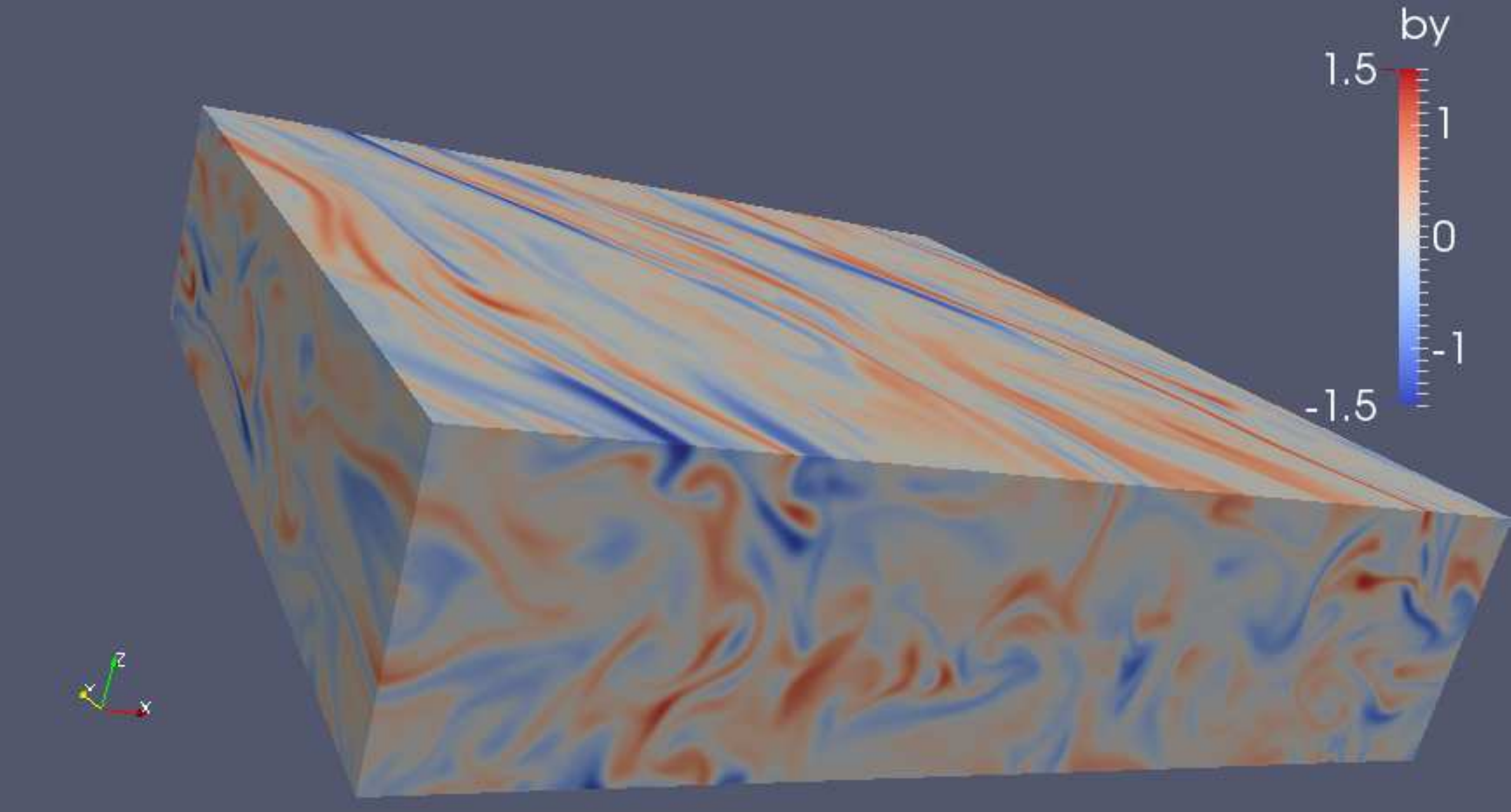}
  \includegraphics[width=\columnwidth]{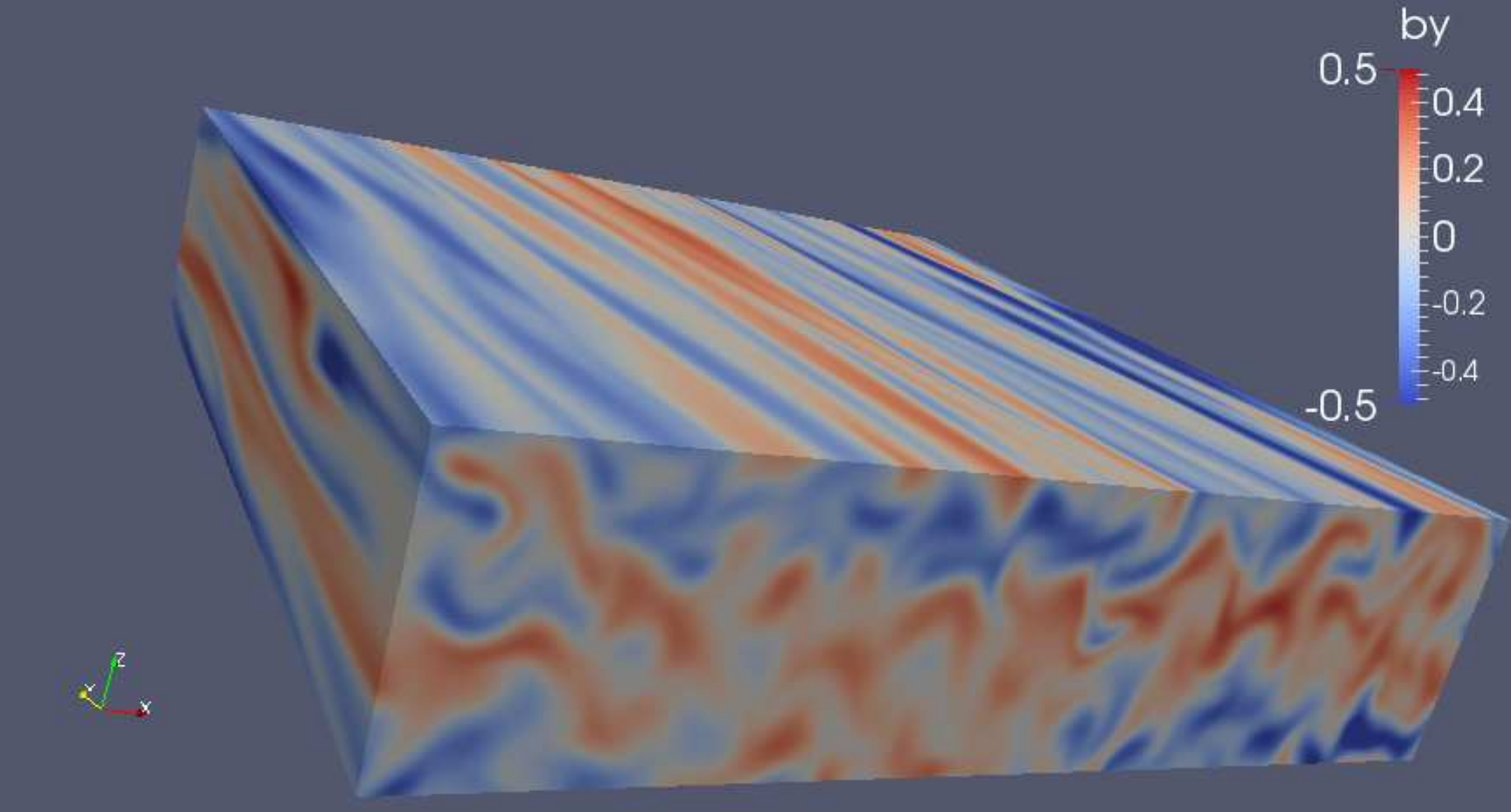}
  \includegraphics[width=\columnwidth]{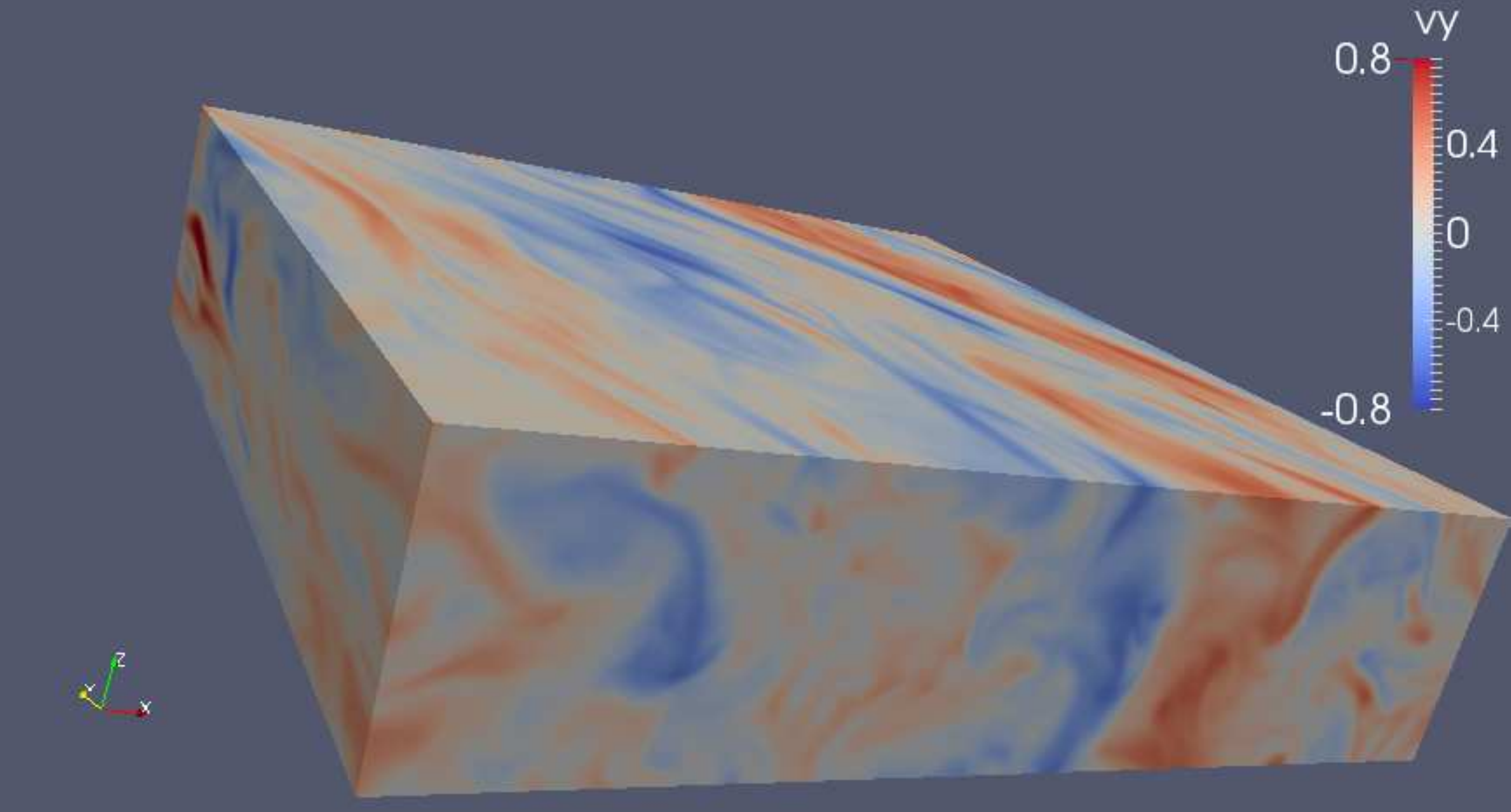}
  \includegraphics[width=\columnwidth]{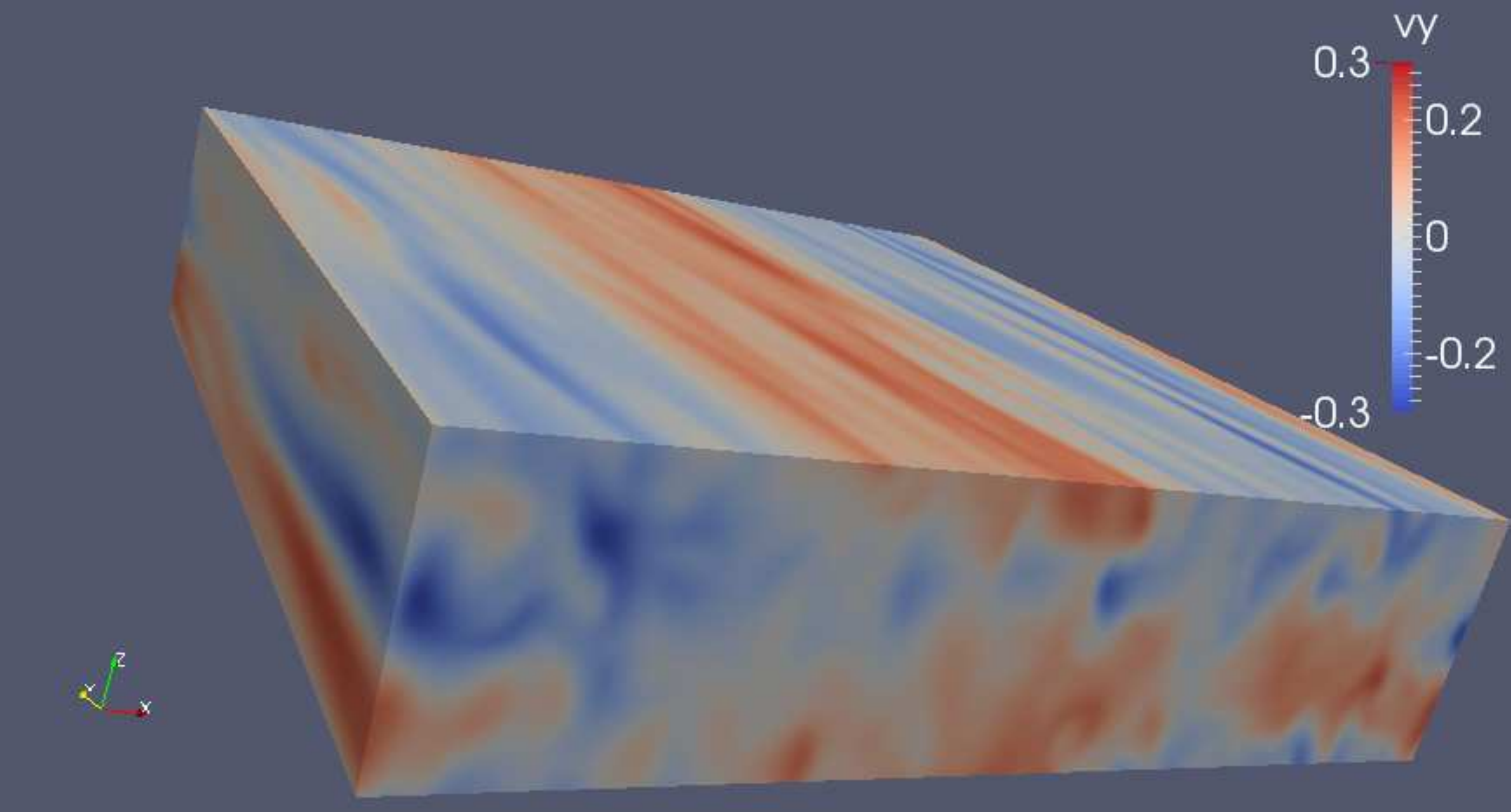}
  \includegraphics[width=\columnwidth]{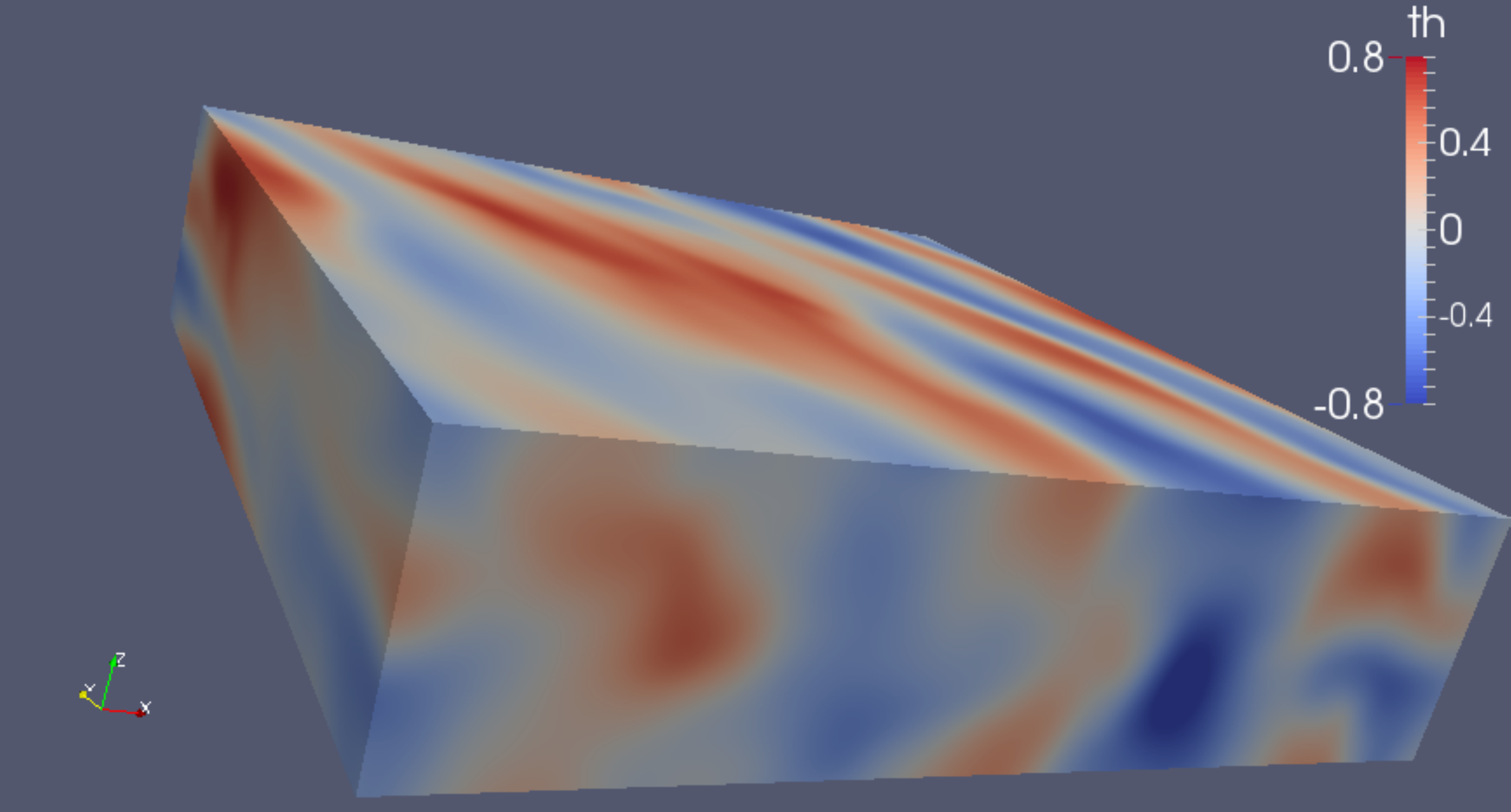}
  \includegraphics[width=\columnwidth]{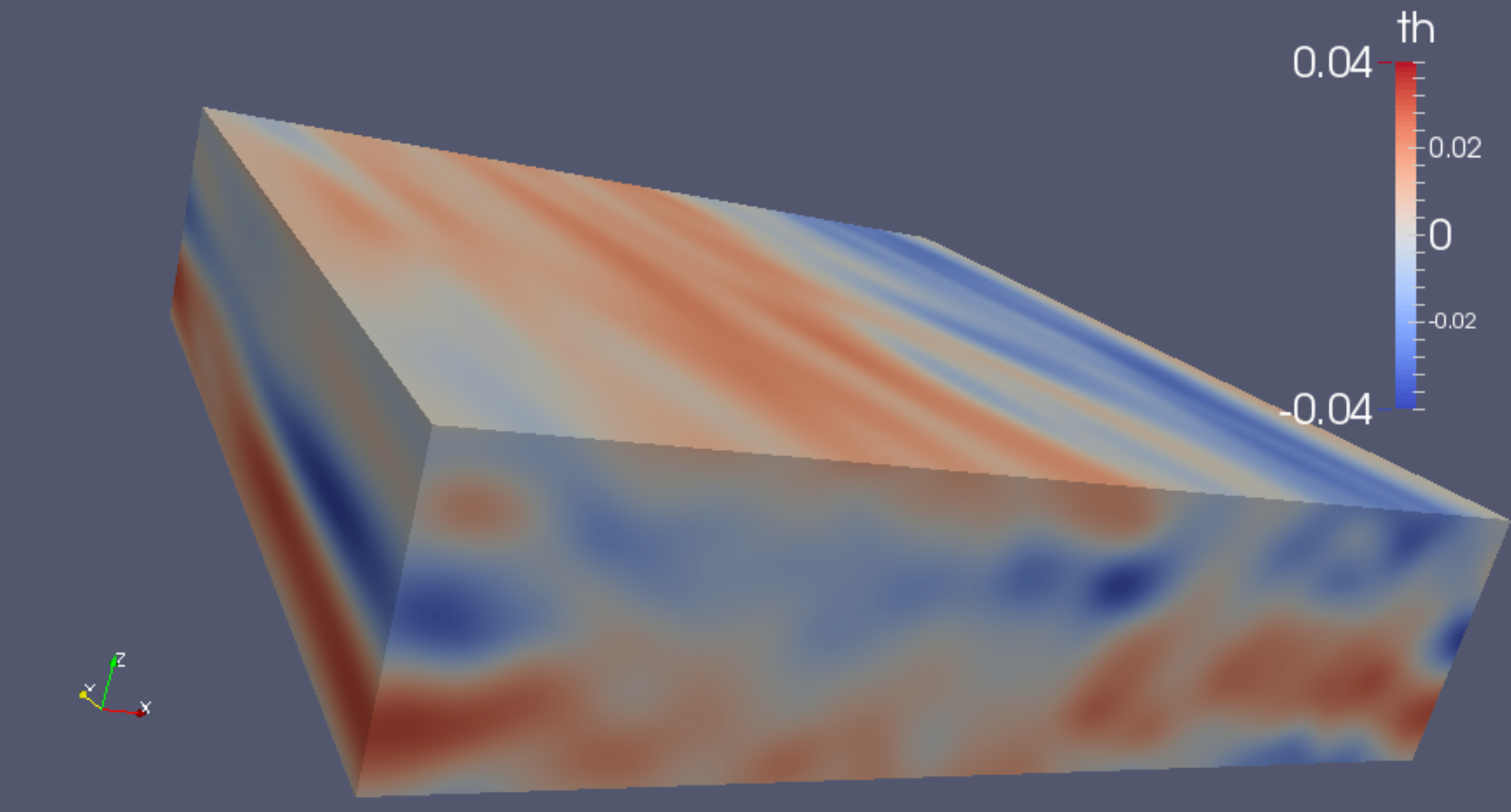}
  \caption{Three-dimensional snapshots at time $t=553\,{\rm ms}$ of two
    simulations with $P_e=100$ for a buoyantly unstable flow with
    $N^2/\Omega^2=-1$ (left column) and buoyantly stable flow with
    $N^2/\Omega^2=10$ (right column). The three rows show the azimuthal
    magnetic field (top), azimuthal velocity (middle), and buoyancy
    variable $\theta$ (bottom). Note that we adjusted the colour scales
    to the level of turbulence, \ie they are different for the two
    columns.}
  \label{fig:images_Pe100}%
\end{figure*}

\begin{figure}
  \centering
  \includegraphics[width=\columnwidth]{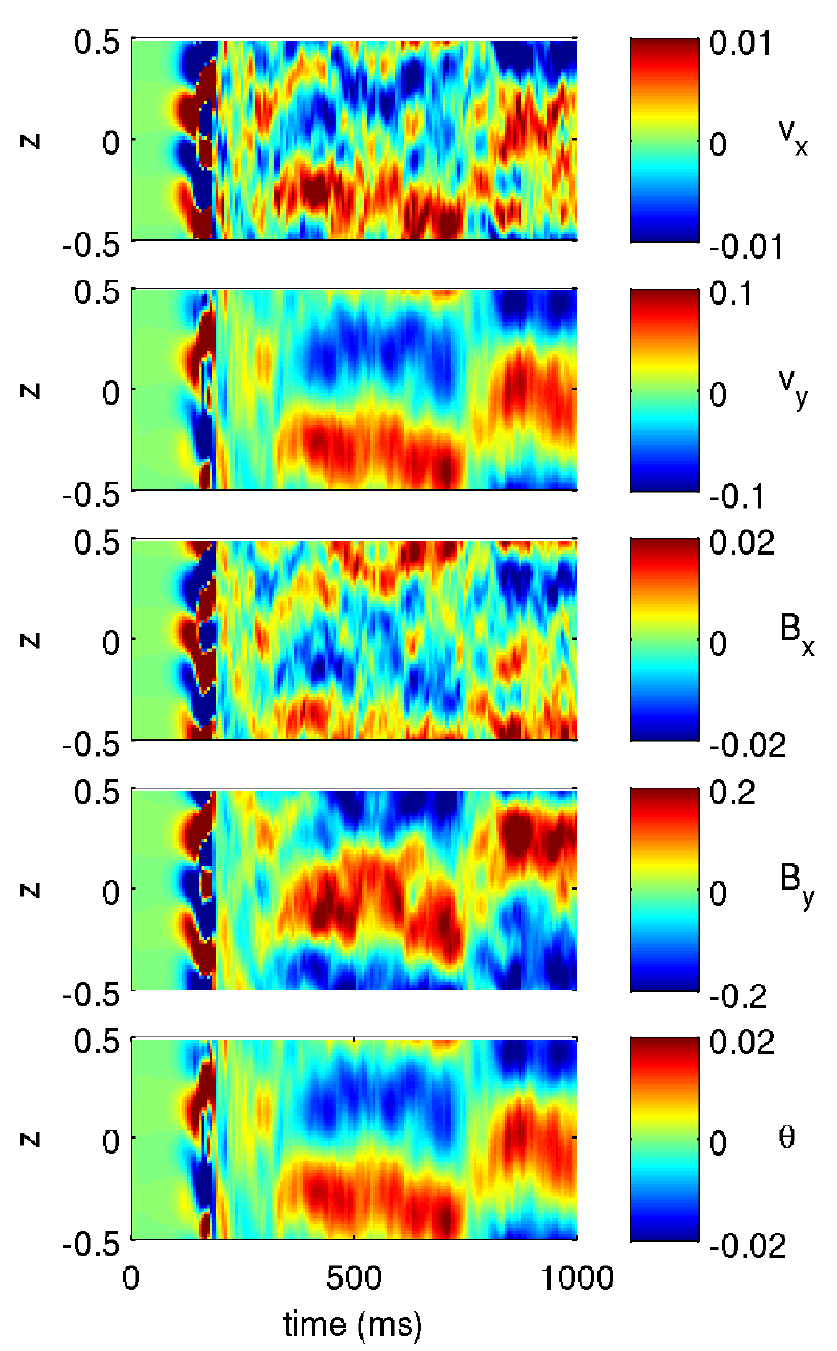}
  \includegraphics[width=\columnwidth]{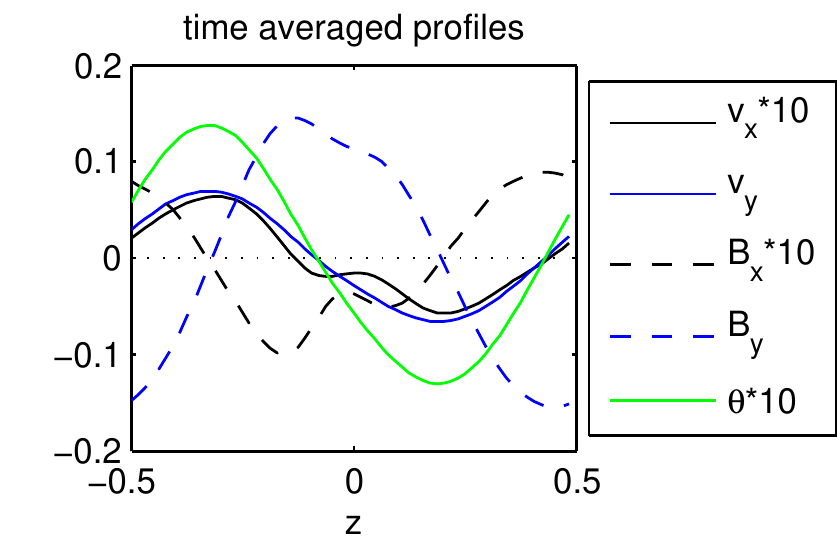}
  \caption{Quasi-stationary channel flow for $N^2/\Omega^2=10$ and
    high thermal diffusion $P_e=100$. The upper five panels show space
    time diagrams (from top to bottom) of the horizontally averaged
    radial velocity, azimuthal velocity, radial magnetic field,
    azimuthal magnetic field, and buoyancy variable $\theta$. The
    bottom panel gives the vertical profiles of these quantities
    averaged over the time interval
    $400\,{\rm ms} \leq t \le 700\,{\rm ms}$. }
  \label{fig:space-time_channels_Pe100}%
\end{figure}

Figure~\ref{fig:images_Pe100} shows snapshots of the azimuthal
magnetic field, the azimuthal velocity, and the buoyancy variable for
two simulations at high thermal diffusion for a flow unstable
($N^2/\Omega^2=-1$; left column) and stable ($N^2/\Omega^2=10$; right
column) to buoyancy. In the latter case, the flow is much more
axisymmetric, and we observe a channel flow structure superposed to a
disordered flow.

The channel flow structure and time-evolution can be made more
apparent by calculating horizontal averages of the flow variables.  We
show space-time diagrams of these averages in
Fig.~\ref{fig:space-time_channels_Pe100} for the simulation with
$N^2/\Omega^2 = 10$ and $P_e=100$. A first transient phase of strong
channel flow occurs between $t\sim100\,{\rm ms}$ and $200\,{\rm ms}$
corresponding to the MRI channels of the exponential growth phase
described in Section~\ref{sec:linear}. After the disruption of these
channels by parasitic instabilities and the onset of turbulence, one
observes no clear channel flow until $t\sim300\,{\rm ms}$. At that
moment, another channel flow appears whose amplitude and phase is
remarkably steady until the end of the simulation, except for a
phase shift that takes place around $t\sim750\,{\rm ms}$. To our
knowledge, this is the first time that such a stationary channel flow
has been seen to appear in MRI turbulence (only recurrent channel
flows were sometimes observed as will be discussed in
Section~\ref{sec:recurrent_channels}).

Let us now compare the structure of this stationary channel flow to
that of growing linear MRI modes. From the space-time diagrams
(Fig.~\ref{fig:space-time_channels_Pe100}) it is already clear that
the wavelength of the stationary flow (one wavelength fits into the
box) is longer than that of the exponentially growing mode (a mixture
of modes with $n=2$ and $n=3$ wavelengths fit into the box), the
latter one corresponding to the linear prediction
\footnote{The linear analysis predicts that the $n=3$ mode is the
  fastest growing mode ($\sigma/\Omega = 0.066$), while the $n=2$ mode
  is the second fastest growing one with a growth rate only slightly
  smaller ($\sigma/\Omega =0.061$). Two further modes ($n=1, 4$) are
  unstable with a roughly twice smaller growth rate.}.
Obviously, the quasi-stationary channel flow does not correspond to
the most unstable MRI mode. The time-averaged profiles
(Fig.~\ref{fig:space-time_channels_Pe100}; bottom panel) show
nonetheless interesting qualitative similarities with linear MRI
modes. The profiles resemble quite closely that of a sinusoidal
  function whose wavelength is equal to the vertical size of the box,
  except for the profiles of the radial velocity and the magnetic
  field which both show some indication of smaller scale structure on
top of the dominant $n=1$ sinusoid. Interestingly, the radial
velocity, azimuthal velocity, and buoyancy variable are in phase,
while the radial and azimuthal magnetic fields are in anti-phase and
shifted by a quarter of a wavelength with respect to the velocity
profiles. This result is in accordance with the structure of a linear
MRI channel mode, and it suggests a link between the stationary
channel flow and linear MRI modes.

In view of the qualitative similarity but quantitative difference
between the stationary channel flow and the linear MRI modes, we
interpret the stationary channel flow as an MRI channel mode
significantly modified by the concurrent
turbulence, which presumably brings it to marginal stability. Further
studies of the physics of the stationary channel flow and its
interaction with turbulence may shed light on the mechanism
determining the level of MHD turbulence. Using enlarged diffusion
coefficients to roughly describe the impact of turbulence on the
channel flow may provide the basis for modelling MRI saturation and
could allow for a deeper interpretation of the numerical
simulations. One could, for example, speculate that the level of
turbulence adjusts so as to bring the largest scale MRI mode to
marginal stability \citep{ogilvie01, guilet12b, guilet13}.

\subsubsection{Quasi-stationary zonal flows}
\label{sec:steady_zonal}

\begin{figure}
  \centering
  \includegraphics[width=\columnwidth]{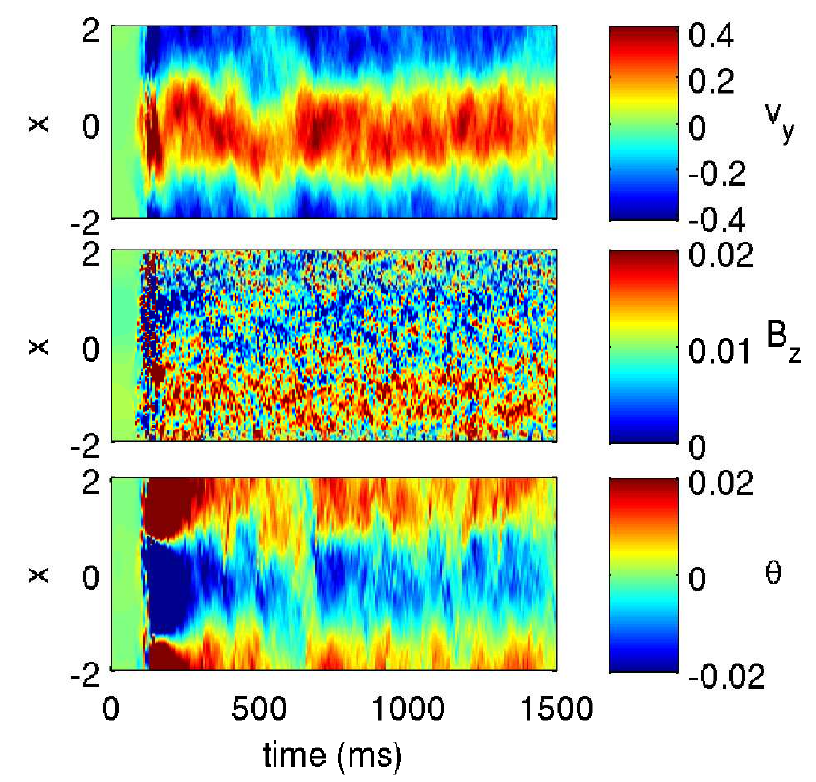}
  \caption{Space-time diagrams for a zonal flow with $N^2/\Omega^2=4$
    and $P_e=100$. The three panels show the azimuthal velocity (top), the vertical magnetic field (middle), and the
    buoyancy variable $\theta$ (bottom). These quantities are averaged
    in azimuthal (y) and vertical (z) direction, and plotted as a
    function of time and radius (x).}
  \label{fig:spatiotemp_zonal}%
\end{figure}

Zonal flows can conveniently be visualised with a space-time diagram
showing the radial profile of azimuthally and vertically averaged
quantities. Fig.~\ref{fig:spatiotemp_zonal} shows such a diagram for a
simulation with $N^2/\Omega^2=4$ and $P_e=100$, where a very stable
quasi-stationary zonal flow is clearly visible. It shows that
  azimuthal velocity has by far the largest amplitude
  amounting to $\sim90\%$ of the kinetic energy, but a modulation with
  a wavelength equal to the radial size of the box is also
  recognisable in the vertical magnetic field
and the buoyancy variable. The modulation of the vertical magnetic
field is shifted by a quarter of a wavelength with respect to that of
azimuthal velocity (\ie the maximum of the magnetic field corresponds
to a zero of the velocity), in agreement with the results of
\citet{bai14c}. We note that the magnetic field zonal modulation
represents a negligible fraction of the total magnetic energy (see
Fig.~\ref{fig:channels_fraction}), but it amounts to a significant
fraction of the mean vertical magnetic field. Thus, it may play a
significant role in the flow dynamics.

The zonal flow shown in Fig.~\ref{fig:spatiotemp_zonal} is
representative of flows for Brunt-V\"ais\"al\"a frequencies
$3\lesssim N^2/\Omega^2 \lesssim 6$. In this parameter regime, the
zonal flow dominates the kinetic energy, and it is very stable and
quasi-steady. At lower and higher values of $N^2$, zonal flows tend to
be more noisy and less stable in that their phase changes somewhat
chaotically with time on timescales of ten or a few tens of orbits.

\subsubsection{Recurrent channel and zonal flows}
\label{sec:recurrent_channels}

\begin{figure}
  \centering
  \includegraphics[width=\columnwidth]{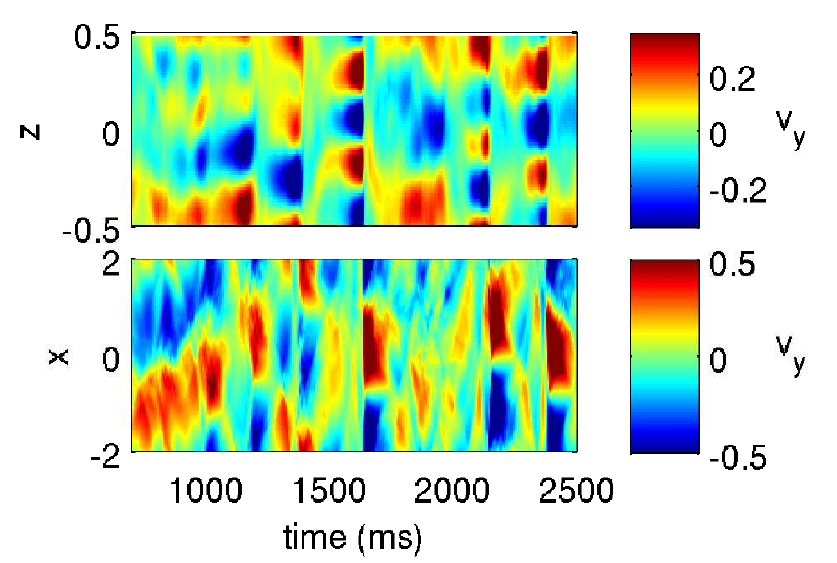}
  \caption{Space-time diagrams of the azimuthal velocity for $N^2/\Omega^2=4$ and $P_e=1000$.  The
    upper panel shows horizontally averaged vertical profiles, where
    recurrent channel flows are clearly visible. The lower panel shows
    vertically and azimuthally averaged radial profiles, where
    transient zonal flows can be recognised just after the disruption
    of the channel flows.}
  \label{fig:space-time_recurrent_channels}%
\end{figure}

\begin{figure*}
  \centering
  \includegraphics[width=\columnwidth]{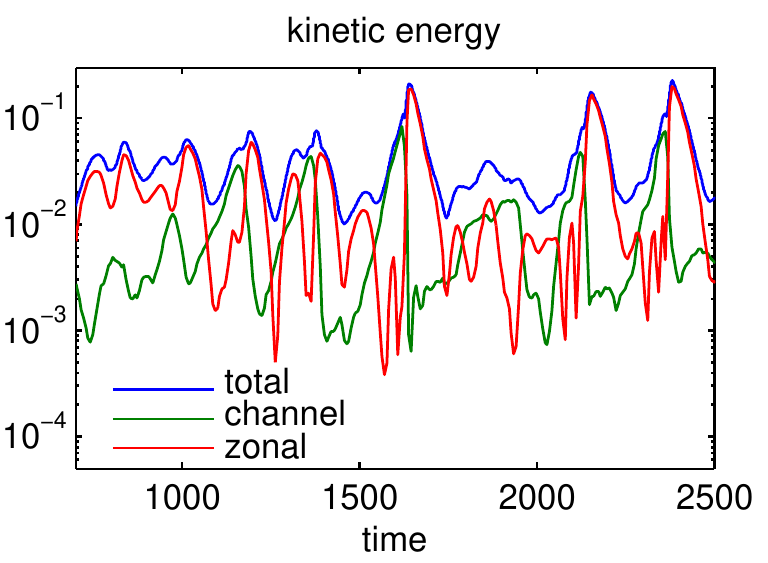}
  \includegraphics[width=\columnwidth]{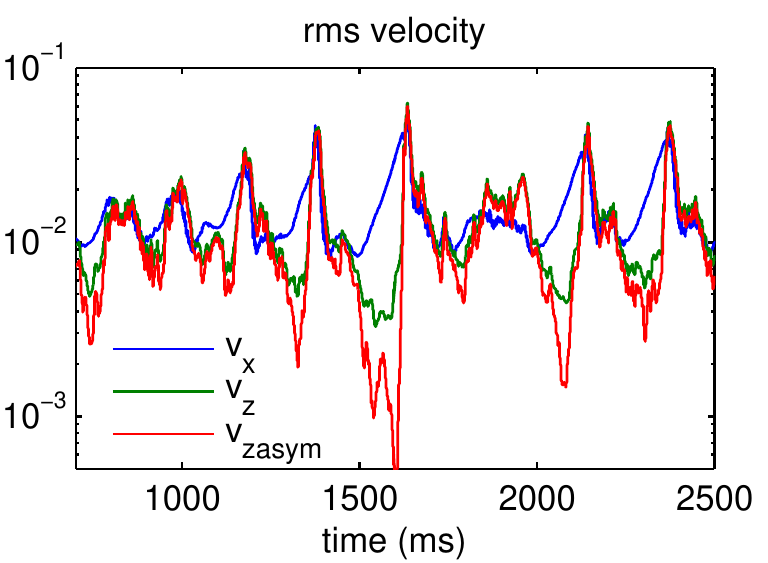}
  \caption{Time evolution of various box averaged quantities showing
    the recurrent growth of channel and zonal flows for a simulation
    with $N^2/\Omega^2 = 4$ and $P_e=1000$. Left panel: time evolution
    of the kinetic energy (blue), the kinetic energy of the channel
    flow (green), and the kinetic energy of the zonal flow
    (red). Right panel: time evolution of the rms averages of the
    radial velocity (blue), vertical velocity (green), and the
    non-axisymmetric component of the vertical velocity (red).}
  \label{fig:recurrent_channels_time}%
\end{figure*}

\begin{figure}
  \centering
  \includegraphics[width=\columnwidth]{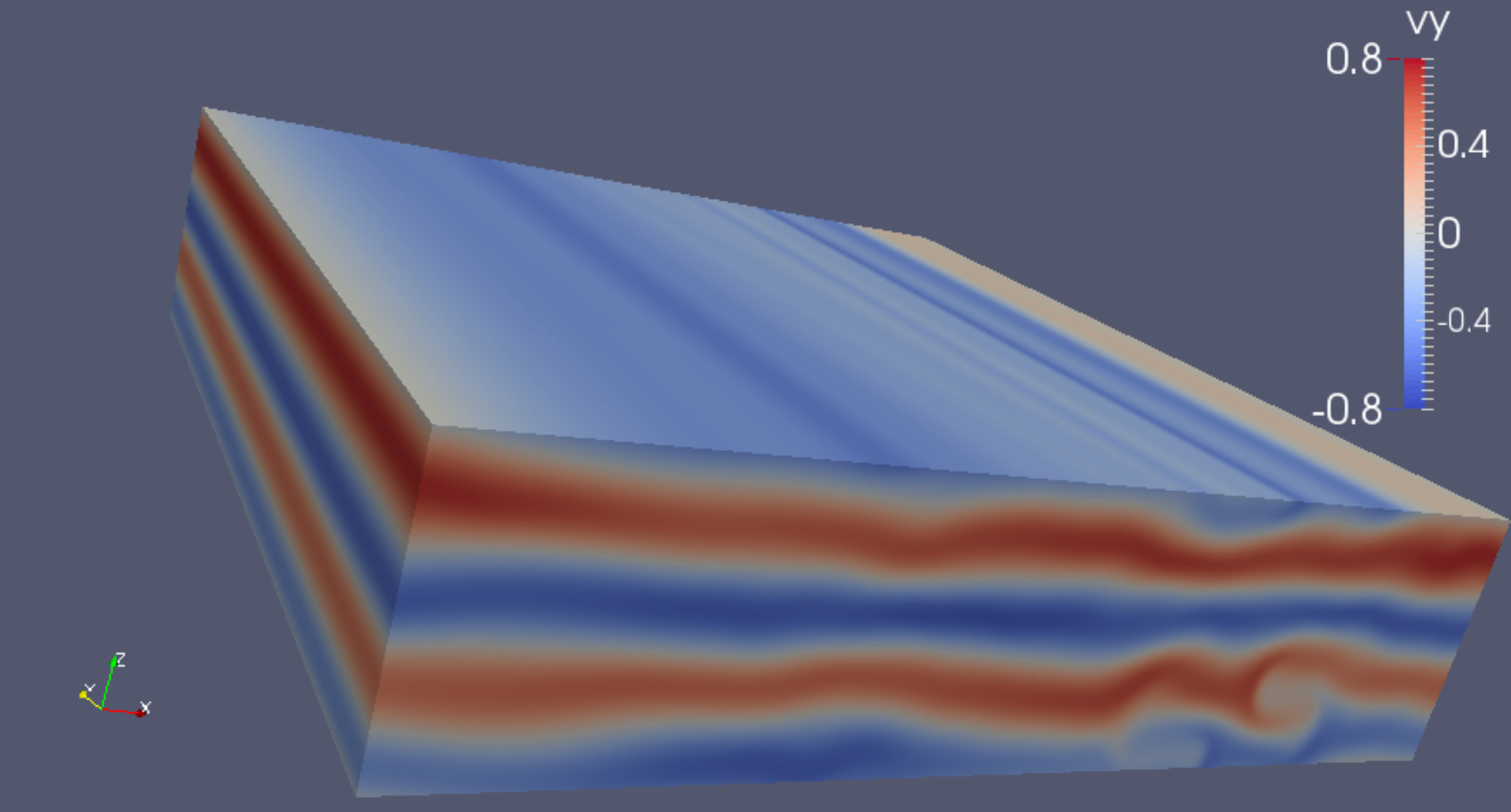}
  \includegraphics[width=\columnwidth]{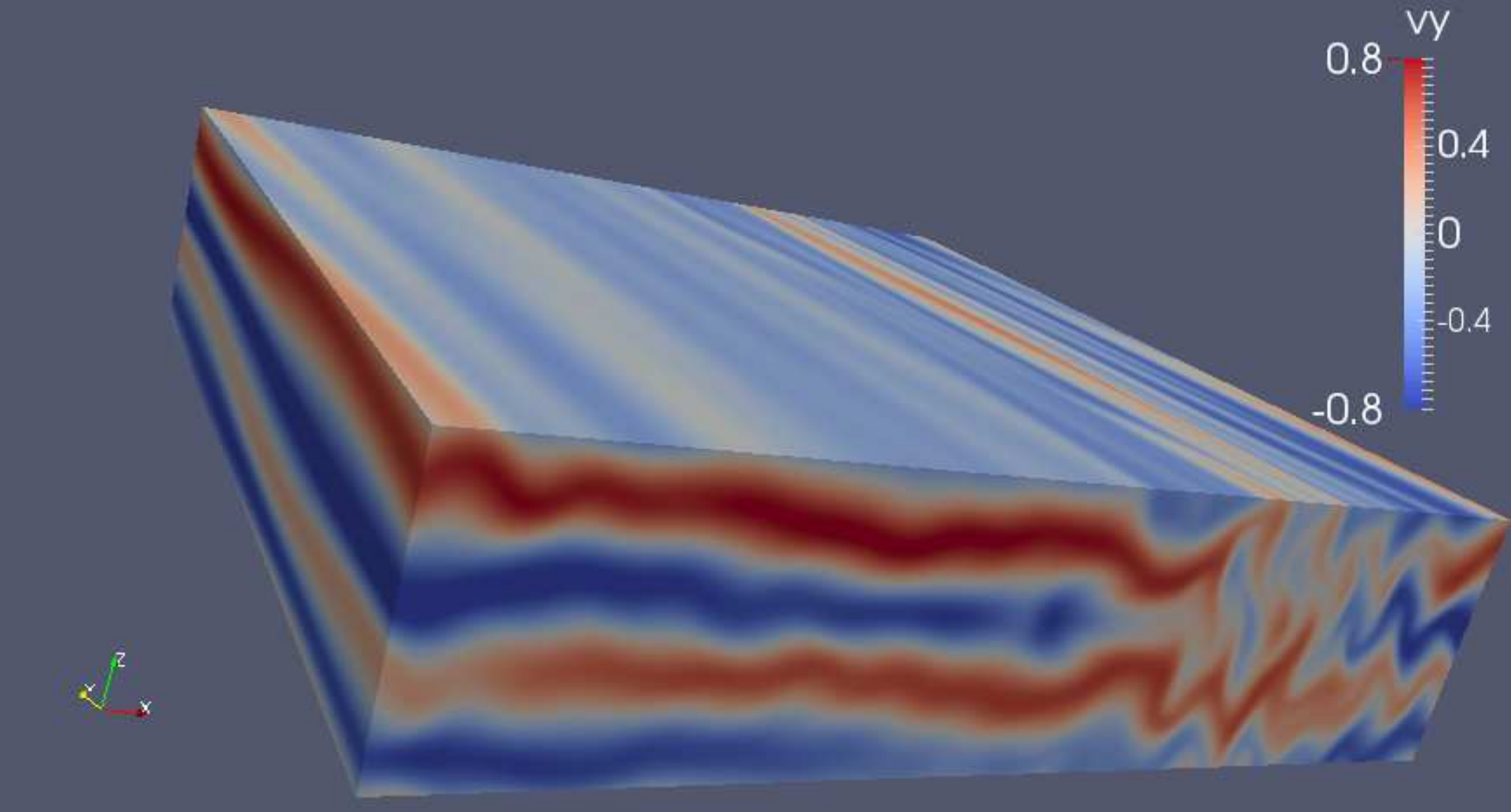}
  \includegraphics[width=\columnwidth]{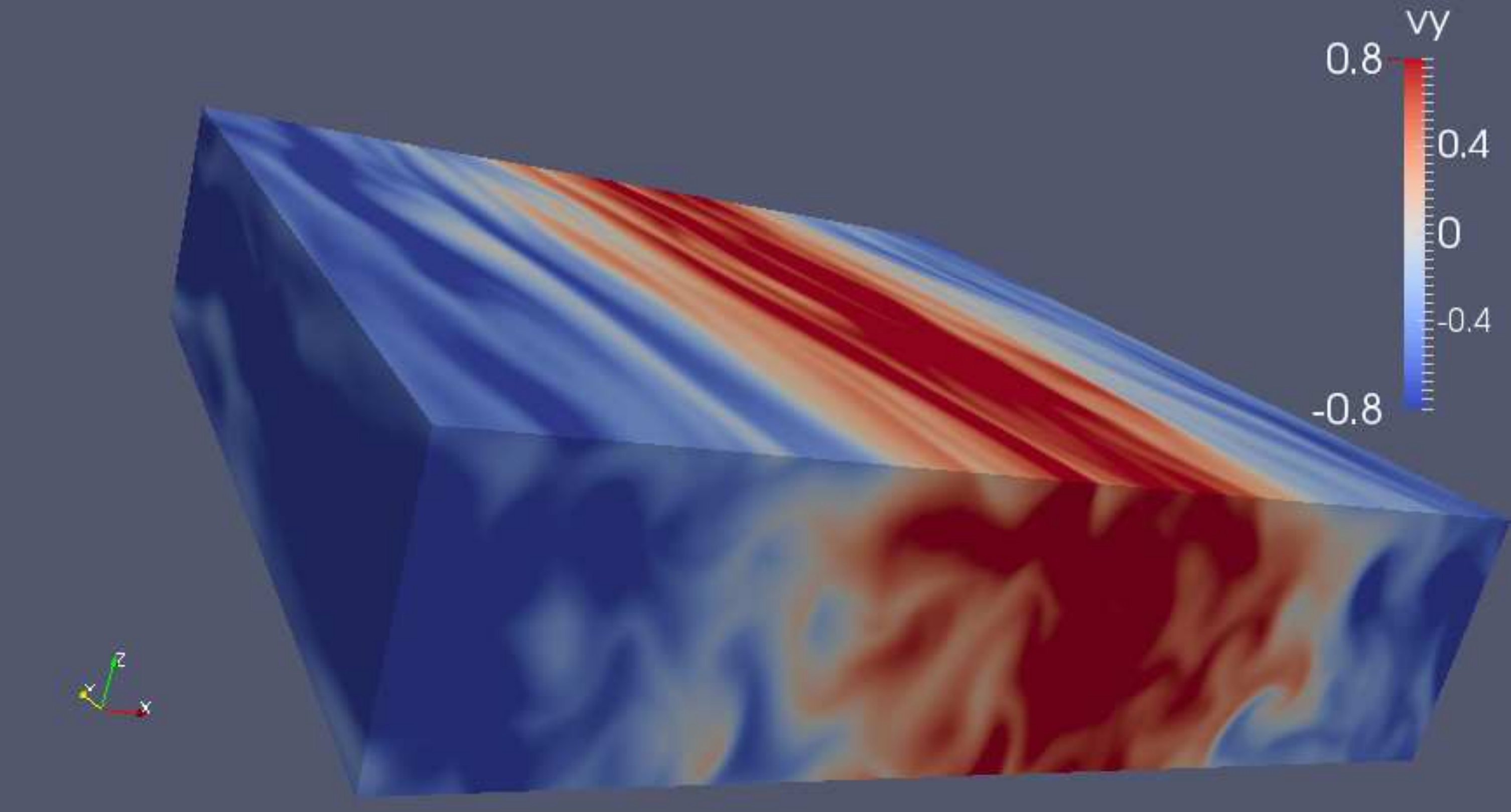}
  \caption{Three snapshots of the azimuthal velocity illustrating the time evolution of recurrent
    channel and zonal flows for a simulation with $N^2/\Omega^2=4$ and
    $P_e=1000$: exponential growth of a channel mode (upper panel),
    its disruption (middle panel), and the formation of a zonal flow
    after the disruption of the channel mode (bottom panel).  The
    snapshots are separated in time by two orbital periods, and they
    correspond (from top to bottom) to times $t= 1615$, $1627$, and
    $1640\,{\rm ms}$, respectively.}
  \label{fig:recurrent_channels_images}%
\end{figure}

At low thermal diffusion and in the stable buoyancy regime, channel
and zonal flows are present too. However, contrary to the high
diffusion case, they are not quasi-stationary. This is illustrated in
Fig.~\ref{fig:space-time_recurrent_channels} with space-time diagrams
of the azimuthal velocity for
$N^2/\Omega^2=4$ and $P_e=1000$. The vertical profile of the
horizontally averaged azimuthal velocity,
$\langle v_y \rangle_{x,y}(z)$ (top panel), shows recurrent phases of
channel mode growth and sudden disruption. The radial profile of the
vertically and azimuthally averaged azimuthal velocity,
$\langle v_y \rangle_{y,z}(x)$ (middle panel) shows that the
disruption of the channel flow is accompanied by the formation of a
strong zonal flow. This zonal flow then decays and another phase of
channel flow growth begins. These successive phases of channel flow
growth, disruption, and zonal flow formation can also be seen in the
left panel of Fig.~\ref{fig:recurrent_channels_time}, which
demonstrates that the formation of the zonal flow exactly coincides
with the disruption of the channel flow. The right panel shows that
during the phase of exponential growth of the channel flow, the radial
velocity (as well as the even higher azimuthal velocity (not shown))
is much larger than the vertical velocity, consistent with a channel
mode dominating the flow. Furthermore, the flow is mostly
axisymmetric, as can be deduced from the fact that the
non-axisymmetric part of the vertical velocity is significantly
smaller than the total vertical velocity. The growth of
non-axisymmetric parasitic instabilities disrupting the channel mode
is visible in the sudden rise of the vertical velocity, and in
particular of its non-axisymmetric part. The disruption of the channel
flow is then followed by a phase of decaying non-axisymmetric
turbulence, during which the radial velocity, the vertical velocity
and the non-axisymmetric part of the vertical velocity have similar
magnitudes.

Recurrent channel flow growth and disruption by parasitic
instabilities has already been observed in numerical simulations of
the MRI in the absence of buoyancy \citep[e.g.][]{lesur07}.
\citet{lesur07} argued that this behaviour is characteristic of the
MRI growing close to its threshold of marginal stability, be it
because the magnetic field is very strong, or the resistivity and
viscosity are very high. This interpretation is consistent with the
results of our simulations, where the MRI is brought close to marginal
stability by the effect of buoyancy.

Figure~\ref{fig:recurrent_channels_images} illustrates the phases of
channel flow (top panel), its disruption (middle panel), and the
formation of a zonal flow (bottom panel). The upper panel shows an
almost pure horizontally uniform channel flow, with only slight
perturbations to the right hand side of the box. Two orbits later,
these perturbations have grown due to parasitic instabilities, and the
channel flow is being disrupted first on the right hand side of the
box, and afterwards in the rest of the box, such that two orbits later
the channel flow has completely disappeared and a zonal flow has
formed. We attribute the formation of the zonal flow to the
non-uniformity of the channel flow disruption over the radial extent
of the box. Indeed, as the disruption of the channel flow happens
first on the right hand side of the box ($1\lesssim x \lesssim 2$), the Maxwell and Reynolds stresses
are first quenched in that part of the box, while they remain high in
the rest of the box. Before the disruption extends to the rest of the
box, the angular momentum flux is therefore non-uniform and the
redistribution of angular momentum creates the zonal flow. To our
knowledge, such episodic zonal flow formation due to the disruption of
a channel flow has never been reported before. Since the mechanism we
propose does not seem directly related to the impact of buoyancy
(whose role is mostly to bring the MRI close to marginal stability),
we suggest that it may also take place in the classical MRI without
buoyancy.

The dynamics described in this section for $N^2/\Omega^2=4$ is
characteristic of simulations with $N^2/\Omega^2 \gtrsim 3$. In this
regime, as $N^2$ increases the typical timescale between phases of
channel flow growth becomes longer. This can be interpreted by the
smaller growth rate of the linear channel modes (see
Fig.~\ref{fig:growth_rate}), which therefore take a longer time to
grow again after their disruption. Note that for
$N^2/\Omega^2 \simeq 1-2$ and $P_e=1000$, the zonal and channel flows
are more similar to the case of high thermal diffusion:
quasi-stationary, though somewhat more variable than those shown in
Sections~\ref{sec:steady_channels} and \ref{sec:steady_zonal}.

\subsection{Time-averaged properties}
\label{sec:time_average}

To characterize the properties of the non-linear state and its
dependence on $N^2$, we performed time and box averages of the
magnetic, kinetic and thermal energies, the energy injection rates,
and the rms values of the three spatial components of velocity and
magnetic field. For each simulation, we adapted the time at which the
averaging is started to avoid the initial transient dynamics (the
first exponential channel growth phase, sometimes followed by a
transient zonal flow). Because of the slower growth in the stable
buoyancy regime, some simulations had to be run for a longer time
period in order to obtain a meaningful average over the non-linear
phase (see Table~\ref{tab:tave}; note that $6.28\,{\rm ms}$
  correspond to one orbit).

\begin{table}
\caption{Time averaging intervals}
\label{tab:tave}
\centering
\begin{tabular} {c|c|c} \hline\hline
$P_e$ & $N^2/\Omega^2$ & interval [msec] \\
\hline
\multicolumn{1}{c}{ 100} & $\leq 4$ & [\phantom{1}  200,\,\phantom{1} 628] \\
\cline{2-3}              &  6       & [\phantom{1}  400,\,           1000] \\
\cline{2-3}              &  8       & [            1000,\,           2000] \\
\cline{2-3}              & 10       & [\phantom{1}  400,\,           1000] \\
\hline
\multicolumn{1}{c}{1000} & $\leq 0.5$ & [\phantom{1} 200,\,\phantom{1} 628] \\
\cline{2-3}              & 1          & [\phantom{1} 300,\, 1600] \\
\cline{2-3}              & 2          & [		1000,\, 3000] \\
\cline{2-3}              & 3          & [           1000,\, 3000] \\
\cline{2-3}              & 4          & [           1000,\, 3000] \\
\cline{2-3}              & 6          & [           2000,\, 6000] \\
\cline{2-3}              & 8          & [           3000,\,10000] \\
\hline
\end{tabular}
\end{table}

\begin{figure*}
  \centering
  \includegraphics[width=\columnwidth]{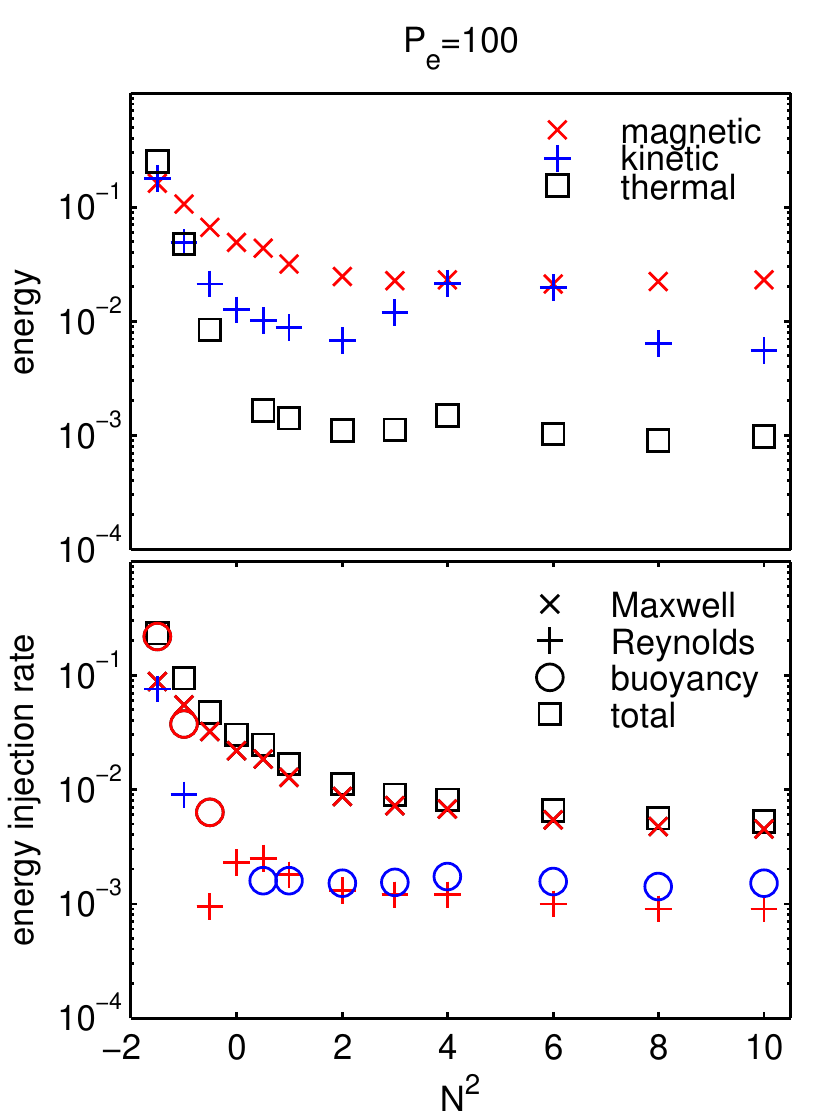}
  \includegraphics[width=0.985\columnwidth]{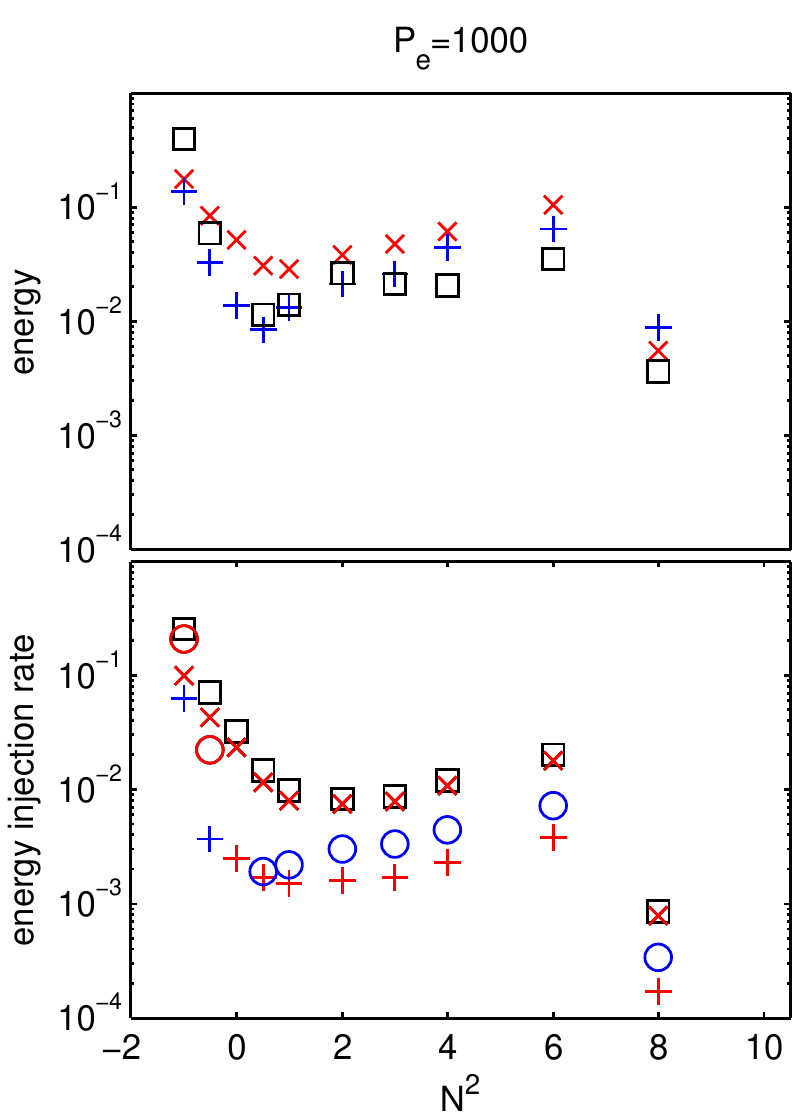}
  \caption{Time and volume averages of the turbulent energy densities and
    energy density injection rates as a function of the Brunt-V\"ais\"al\"a
    frequency. Panels in the left and right column correspond to high
    ($P_e=100$) and low ($P_e=1000$) thermal diffusion, respectively.
    They show the magnitude of magnetic (red $\times$), kinetic (blue $+$), and thermal (black squares) energies in the top
    panels, and energy injection rates due to Maxwell stress ($\times$
    symbols), Reynolds stress ($+$ signs), buoyancy force work
    (circles), and sum of the three (squares) in the bottom
    panels. Positive injection rates are shown in red colour, while
    negative ones (\ie energy is removed from turbulent motions) are
    shown in blue.}
  \label{fig:energy_stress_vs_BV}
\end{figure*}

Figure~\ref{fig:energy_stress_vs_BV} illustrates how the magnetic,
kinetic, and thermal energies, and the energy injection rates depend
on the Brunt-V\"ais\"al\"a frequency. For high thermal diffusion,
magnetic energy and Maxwell stress decrease monotonically with $N^2$,
as one may intuitively expect. Importantly, the dependence gets much
weaker as $N^2$ increases, to the extent that the magnetic energy is
 practically independent of $N^2$ for $N^2/\Omega^2 \geqslant 3$,
while the Maxwell stress very slowly declines in this regime. As
a result, the magnetic energy is more affected by an unstable
buoyancy, than by a stable one. It increases by a factor of $3.3$
between $N^2=0$ and $N^2/\Omega^2=-1.5$, but it decreases only by a
factor $2.2$ between $N^2=0$ and $N^2/\Omega^2=10$. In the stable
buoyancy regime ($N^2>0$), the thermal energy and the
  corresponding injection rate are almost
independent of $N^2$, and both are much smaller than the
  corresponding magnetic quantities. By contrast, in the unstable
buoyancy regime, they increase rapidly in magnitude as $N^2$, up to the point where the thermal energy and the
corresponding injection rate are the dominant contributions at
$N^2/\Omega^2=-1.5$, suggesting that turbulence is driven then more by
buoyancy than by the shear.  Note that the increased turbulence strength in the unstable buoyancy regime is similar to that observed in the context of accretion discs by \citet{hirose14} and \citet{hirose15}. The geometry is nonetheless different: in accretion discs the buoyancy force is mostly vertical, while in our setup it is radial.

At low thermal diffusion, we found qualitatively similar trends for
$N^2/\Omega^2 \lesssim 1$, but with a somewhat steeper dependence on
$N^2$, a result which agrees with the idea that a lower thermal
diffusion allows buoyancy to act more effectively. The behaviour for
$2-3 \lesssim N^2/\Omega^2\lesssim 6$ is more surprising. All energies
and energy injection rates increase in magnitude with $N^2$, in spite
of the fact that, as in the high diffusion case, buoyancy is actually
removing energy from the flow (because the buoyancy injection rate is
negative). This behaviour is found in the parameter regime showing
strong recurrent channel flows, where the energy is predominantly
contained in axisymmetric structures (see
Section~\ref{sec:coherent_flows}). The increase in energy and energy
injection rate is due to the fact that, at higher $N^2$, recurrent
channel flows reach larger amplitudes before they are disrupted by
parasitic instabilities. The cause of the increase of the channel flow amplitude at termination is unclear, but
may be linked to the so far unknown impact of buoyancy on parasitic
instabilities. In this respect it is interesting to note that, at high
thermal diffusion, the amplitude at which the first phase of channel
mode growth is terminated increases with $N^2$ (as can be seen in
Fig.~\ref{fig:maxwell_vs_time}). This may be linked to the increasing
energy observed in the non-linear phase for low thermal diffusion. It
also highlights the fact that the amplitude at which channel mode
growth is terminated is not necessarily correlated with the strength
of MRI turbulence (which decreases with $N^2$ for high thermal
diffusion).

\begin{figure*}
  \centering
  \includegraphics[width=2\columnwidth]{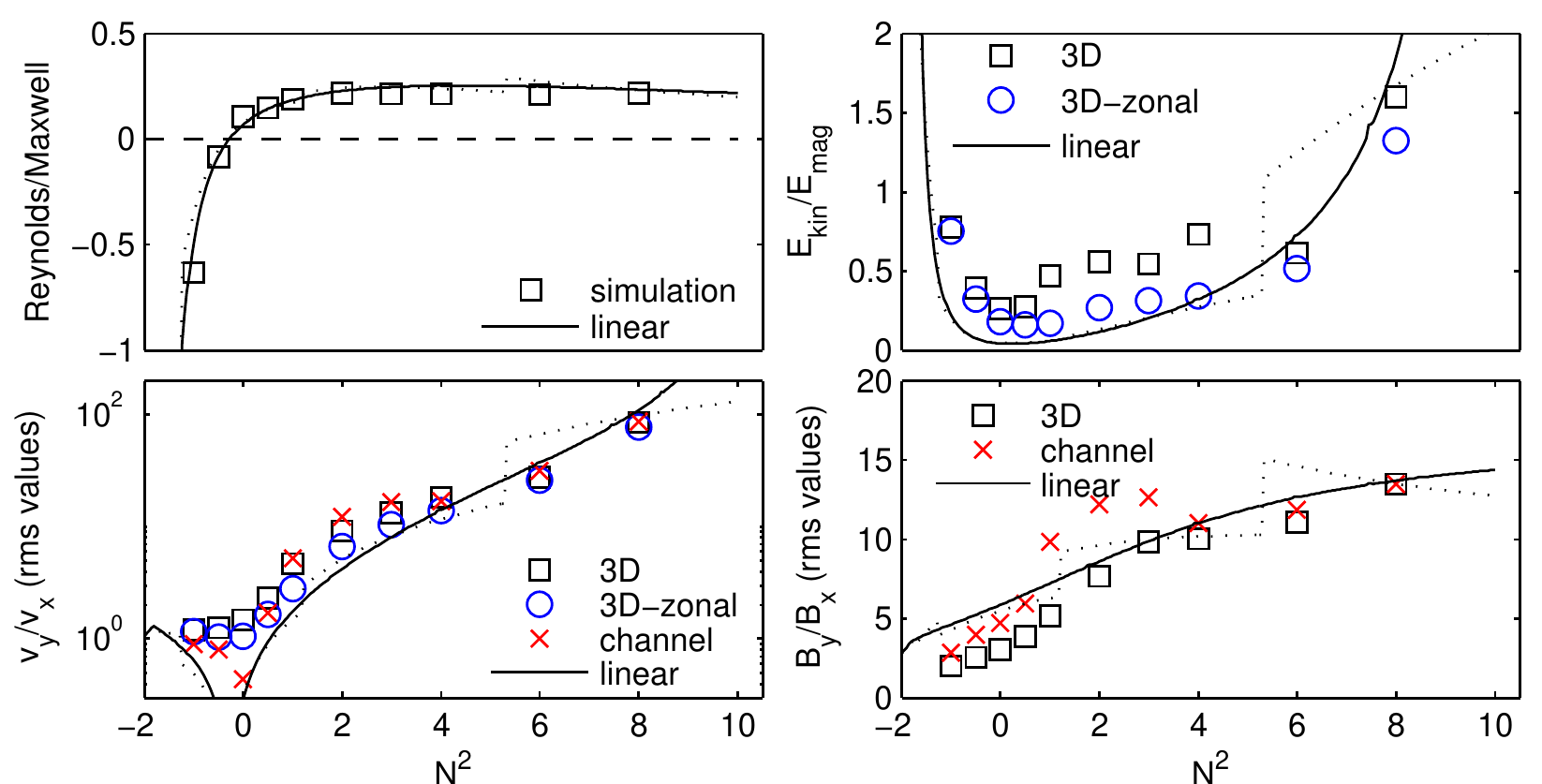}
  \caption{Ratios of different time and volume averaged diagnostics as
    a function of $N^2/\Omega^2$ for low thermal diffusion
    ($Pe=1000$): Reynolds stress to Maxwell stress (top left), kinetic
    energy to magnetic energy (top right), azimuthal velocity to
    radial velocity (rms averages; bottom left), azimuthal magnetic
    field to radial magnetic field (rms averages; bottom right). Black
    squares give the results for the whole flow, blue circles when the
    zonal flow is subtracted, and red crosses for the channel
    flow. The lines show properties of the most unstable linear modes:
    discretised modes (dotted) that fit in the numerical box, and
    continuous modes (solid) relevant to an infinitely high box (which
    may be interpreted crudely as some weighted average of the two
    most unstable discretised modes).}
  \label{fig:ratios_Pe1000}
\end{figure*}

\begin{figure*}
  \centering
  \includegraphics[width=2\columnwidth]{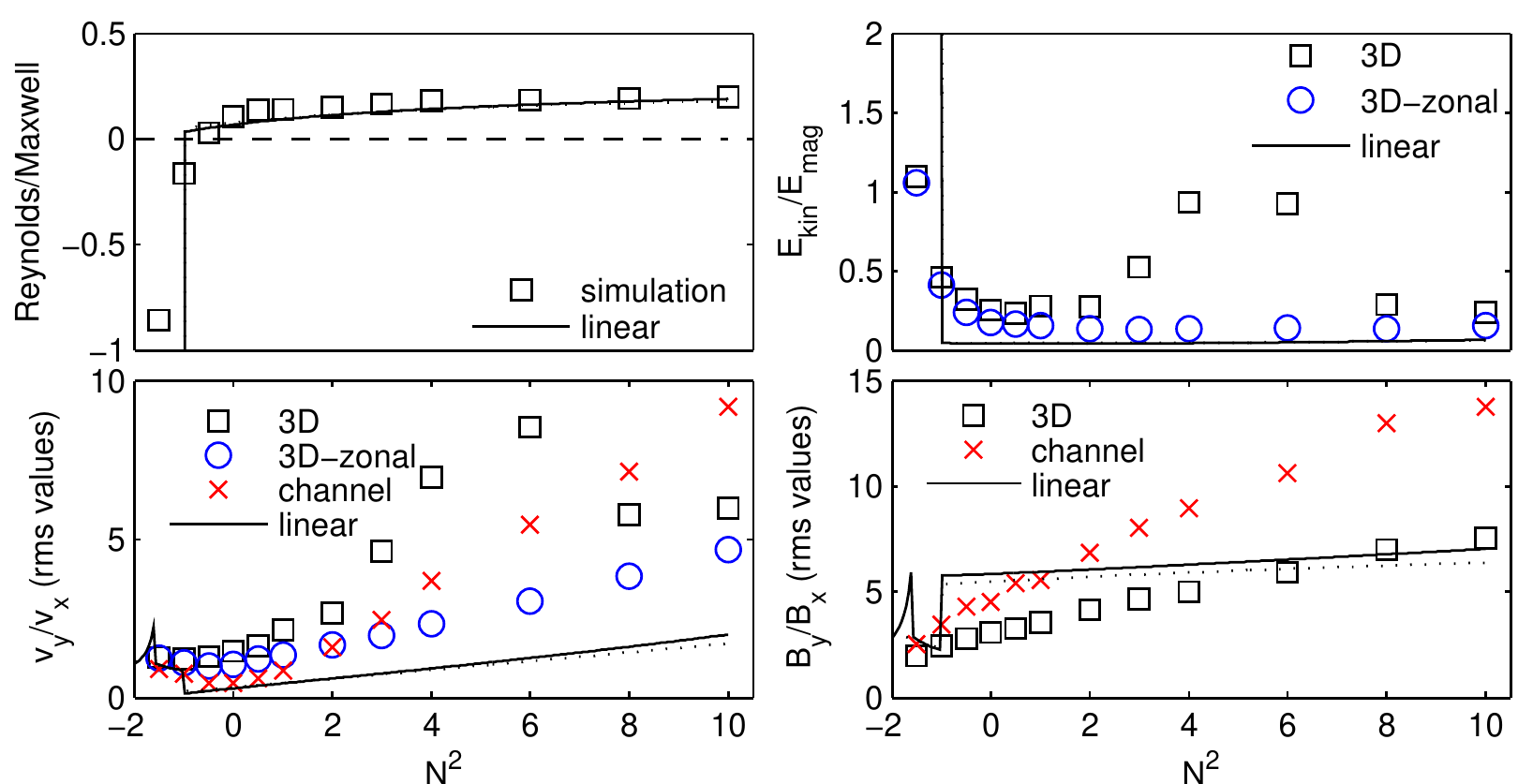}
  \caption{Same as Fig.~\ref{fig:ratios_Pe1000}, but high thermal
    diffusion ($P_e=100$).}
  \label{fig:ratios_Pe100}
\end{figure*}

Figures~\ref{fig:ratios_Pe1000} and \ref{fig:ratios_Pe100} show the
dependence of the ratio of different flow quantities on $N^2$:
Reynolds stress to Maxwell stress (top left), kinetic energy to
magnetic energy (top right), radial to azimuthal velocity amplitudes
(bottom left), and radial to azimuthal magnetic field amplitudes
(bottom right). An advantage of studying such ratios is that they
allow for a comparison of the flow properties of the non-linear phase
to that of linear channel flows \citep{pessah06a}. Indeed, while the
amplitude of a quantity (say the Reynolds stress) due to a channel
mode is unknown if one does not know a priori the mode amplitude, the
ratio of two quantities (e.g.\ Reynolds stress to Maxwell stress) is
well defined for a channel mode independently of its amplitude (see
Appendix~\ref{sec:linear_analysis}). Figure~\ref{fig:ratios_Pe1000}
shows that, at low thermal diffusion, all these ratios vary
significantly with $N^2$ and are in general agreement with the
prediction from the linear modes. This agreement is not surprising in
the regime of stable buoyancy where channel modes contribute a
significant fraction of the energy, but it is surprising in the regime of
  unstable buoyancy where the fraction of energy contained in channel
modes is insignificant. The (at least qualitative) agreement in the
latter regime therefore suggests that linear dynamics leaves an
imprint on the properties of turbulence even when channel modes do not
dominate the flow.

Interestingly, in the regime of unstable buoyancy ($N^2/\Omega^2=-1$
and $-0.5$) the Reynolds stress changes sign and becomes negative,
while it is always positive (corresponding to an outward transport of
angular momentum) for the MRI in the absence of buoyancy. At
$P_e=1000$, this behaviour is reproduced amazingly well by the linear
analysis. When the Reynolds stress becomes negative, the Coriolis
force is not anymore a driver of the radial motion in the channel mode
contrary to the classical MRI. Then the buoyancy force drives the
  radial motion instead. The magnetic field nevertheless still plays
a crucial role in the instability mechanism by reducing the
stabilising effect of the Coriolis force, because no non-oscillatory
growth is possible in this regime in the absence of a magnetic field
(see Section~\ref{sec:linear}). Hence, we interpret the
  instability in the unstable buoyancy regime as a mixed
  magneto-rotational buoyant instability, and the increasing ratio of
  kinetic to magnetic energy by the increasing role played by the
buoyancy force compared to the magnetic one.

Another interesting trend visible in Fig.~\ref{fig:ratios_Pe1000}
  concerns both the ratio of azimuthal to radial velocity and the
  ratio of azimuthal to radial magnetic field. These ratios increase
tremendously with $N^2$, reaching a value of $100$ for the former
one. This may qualitatively be interpreted by
the fact that for increasing $N^2$, the buoyancy force prevents
  more and more efficiently radial motions. In the regime of stable buoyancy magnetic and kinetic energy are
therefore dominated by the azimuthal component of the magnetic
  field and the velocity, respectively.  The vertical components of
  both quantities (not shown) are of the same order of magnitude as
 their radial components. 

We found qualitatively similar trends in the case of high thermal
diffusion 
\footnote{Except for an increase of the ratio of kinetic to magnetic
  energy at high $N^2$, which is observed for low but not for high
  thermal diffusion.}
(Fig.~\ref{fig:ratios_Pe100}). 
In the regime of unstable buoyancy, the Reynolds stress becomes
negative and the ratio of kinetic to magnetic energy increases. The
ratio of azimuthal to radial velocity and the ratio of azimuthal to
radial magnetic field increase significantly with $N^2$. Contrary to
the low thermal diffusion case, these trends are not reproduced
quantitatively, however, by the linear predictions. Still, it is
interesting that most qualitative trends are shared between the
simulations and the linear modes (e.g. the ratio of azimuthal to
radial velocity increases with $N^2$), suggesting that linear analysis
may still provide a qualitative understanding of these trends.

Finally, Fig.~\ref{fig:ratios_Pe100} provides interesting information
on the coherent channel and zonal flows described in
Section~\ref{sec:coherent_flows}. Our simulations show that the
ratios of azimuthal to radial velocity and of azimuthal to radial
magnetic field of the quasi-stationary channel flow are quantitatively
different from those of the most unstable linear mode (but with a
similar qualitative trend). This confirms our conclusion from
Section~\ref{sec:steady_channels} that quasi-stationary channel flows,
while related to MRI modes, have a different structure presumably
because they are significantly modified by the concurrent
turbulence. The simulations further show that the zonal flows
described in Section~\ref{sec:steady_zonal} are responsible for the
bump in kinetic energy observed in Figs.~\ref{fig:energy_stress_vs_BV}
and \ref{fig:ratios_Pe100} for $3 \lesssim N^2/\Omega^2\lesssim 6$.
Remarkably, at the peak of the zonal flow ($N^2/\Omega^2=4$ and 6) the
kinetic energy is in equipartition with the magnetic energy. Whether
this is a mere coincidence or a general property of zonal flows is
unclear. Interestingly, if we subtract the contribution of the
  zonal flow, the remaining kinetic energy (blue circles in the top
right panel of Fig.~\ref{fig:ratios_Pe100}) has a smooth flat
dependence on $N^2$, with no visible impact of the zonal flow (like
the total magnetic energy). This suggests that the zonal flow does not
significantly impact the level of turbulence developing on top of the zonal flow.

\section{Discussion}
\label{sec:discussion}

In this section, we discuss the appropriateness of some of the
simplifying assumptions made in our study: use of the Boussinesq
approximation (Section~\ref{sec:boussinesq_validity}), the assumption
of a constant shear rate
\footnote{In the shearing box formalism the box integrated shear rate is kept constant, while its
  local value is allowed to vary (which is the case in zonal
    flows).}
(Section~\ref{sec:shear_evolution}), and the local approximation, in
particular whether our results depend on the box size
(Section~\ref{sec:box_size}).

\subsection{Validity of the Boussinesq approximation}
\label{sec:boussinesq_validity}
Because our simulations are the first attempt to apply the Boussinesq
approximation to the problem of MRI growth in a PNS, it
is important to assess to what extent this approximation provides an
appropriate description of the flow dynamics. Several conditions
should be satisfied for the approximation to be justified. Firstly,
the flow velocity and the Alfv\'en velocity should be much smaller
than the sound speed, so that the flow is approximately
incompressible. In all our simulations, this requirement holds very
well, because $v^2/c_s^2 \lesssim v_A^2/c_s^2 \lesssim 10^{-4}$, the
sound speed $c_s \simeq 5\times10^9\,{\rm cm\,s^{-1}}$ being estimated
from the PNS model studied by \citet{guilet15}. Secondly, the
  relative density perturbation associated with buoyancy,
  $\delta \rho/\rho_0 = -\theta N^2/g$, should be small. Using the
gravitational acceleration $g\simeq6\times 10^{13}\,{\rm cm\,s^{-2}}$
\citep[again from the model of][]{guilet15}, we find that in all
  our simulations its rms value is $\delta \rho/\rho_0 \lesssim 1.5\times 10^{-3}$, \ie it is indeed
quite small. Thirdly, the Boussinesq approximation neglects the
  influence of a background density gradient, an assumption that
  is no longer justified when the size of the box approaches the
density scaleheight. For the box size used in our study, the density
at the radial surfaces of the box should actually differ from
the value at the centre of the box by about $30\%$.

The neglect of the background density gradient is by far the main
limitation of the Boussinesq approximation in our simulations. Note
that this limitation is common to all local models, even those not
employing the Boussinesq approximation \citep[e.g.][]{masada12}. It
can, in principle, be lifted only by using a global model, or a
semi-global model \citep{obergaulinger09}. So far, global models
resolving the MRI have only been published for two dimensional flows
\citep{sawai13,sawai14}, which is problematic for a study of MHD
turbulence. On the other hand, the inclusion of the density gradient
in the semi-global models of \citet{obergaulinger09} renders the use
of periodic boundary conditions problematic, and causes important
artefacts at the radial boundaries of the computational domain (e.g.,
a fast flattening of the rotation and entropy profiles, which prevents
a meaningful study of the non-linear phase of the MRI). The radially
global (but vertically local) model of \cite{masada15} may not
suffer from such boundary artefacts, but the horizontal resolution
that they could afford is in our opinion much too coarse for an accurate study of the
MRI. We conclude that so far no satisfying solution has been found to
study the impact of the large-scale density gradient on the MRI, and
that presently the use of local simulations in the Boussinesq
approximation is probably the best choice to study the impact of
buoyancy on the MRI.

\subsection{Energy dissipation rate and timescale of shear profile evolution}
\label{sec:shear_evolution}
Another limitation of local simulations is that they cannot describe
the evolution of the global rotation profile. In shearing box
simulations with shear-periodic boundary conditions in the radial
direction, the shear rate averaged over the box is kept constant. This
is justified only as long as the rotation profile has not been
significantly modified by the angular momentum transport driven by the
MRI.

One can estimate the timescale over which the rotation profile is
significantly changed by the ratio of the total available shear energy
to the rate at which it is extracted by the combined effects of
Reynolds and Maxwell stresses. The available shear energy is estimated
in the fast rotating PNS model studied by \citet{guilet15} by
comparing the rotational energy contained in the differentially
rotating PNS to that of a PNS in solid rotation with the same total
angular momentum. In this way, we estimate a shear energy of
$\sim 4\times10^{50}\,{\rm erg}$.  The specific rate at which energy
is extracted from the shear
\footnote{The total rate of energy injection into turbulence can be
  higher in the regime of unstable buoyancy (up to
  $3\times 10^{18}\,{\rm erg\,s^{-1}\,g^{-1}}$) due to the buoyancy force, but this energy is not extracted from the shear.}
is between $5\times10^{16}\,{\rm erg\,s^{-1}\,g^{-1}}$ in the regime
of stable buoyancy for high thermal diffusion, and
$5\times10^{17}\,{\rm erg\,s^{-1}\,g^{-1}}$ in the unstable buoyancy
regime. If we assume that this specific energy extraction rate is
uniform in the differentially rotating envelope of the PNS (which
contains a mass of $1.6\times10^{33}\,{\rm g}$ in the PNS model
considered), we estimate a total extraction rate of shear energy
between $8\times 10^{49}\,{\rm erg\,s^{-1}}$ and
$8\times 10^{50}\,{\rm erg\,s^{-1}}$. The typical timescale over which
the shear energy is extracted is therefore between $500\,{\rm ms}$ in
the unstable buoyancy regime and $5\,{\rm s}$ in the stable one,
  which is significantly longer than the dynamical timescale
$1/\Omega = 1\,{\rm ms}$, and also in most cases longer than the time
needed to establish the non-linear statistically steady state. This
justifies to some extent the relevance of the local approach, but one
should keep in mind that in some cases the shear profile should
actually have evolved before the end of the simulation.

\subsection{Local approximation : dependence on box size}
\label{sec:box_size}
One problem greatly limiting the predictive power of local numerical
simulations, is that the result is likely to depend on the size of the
computational domain. For the MRI in the absence of buoyancy and in
the presence of a vertical magnetic field, numerical simulations have
indeed found that the level of turbulence as measured by the angular
momentum flux is proportional to the vertical size of the box
\citep{hawley95}. Since the relevant box size to be adopted is a
priori unknown, this renders very uncertain the use of local models to
predict magnetic field strength and angular momentum transport in PNS
(and casts some doubts on the scaling formulae given by
\citet{masada12}, who did not discuss the dependence of their results
on the box size). One may argue the density scale height to be a
natural upper limit to the scales available to MRI driven turbulence,
such that choosing a local box of a size comparable to the density
scale height could be an appropriate choice. Whether this argument
actually holds can be checked only with global numerical simulations
covering at least several density scale heights if not the whole
PNS. While 3D simulations of such global models seem out of reach
presently, investigating the dependence of local models on the domain
size would already shed more light on its potential impact on the
results. The dependence on the box size of MRI simulations that
include buoyancy is so far unknown, and should be explored in the
future.

\section{Conclusions}
\label{sec:conclusion}
We have investigated the impact of buoyancy on the properties of the
MRI in protoneutron stars. For this purpose, we performed numerical
simulations of a local model of a PNS using for the first time the
Boussinesq approximation. This approximation has proven very useful to
take into account the impact of entropy and lepton number gradients
while keeping the advantages of a local model, \ie clean shearing
periodic boundary conditions and low computational cost allowing for
an exploration of parameter space. We compared the results of our
simulations to those of a linear analysis presented in
Appendix~\ref{sec:linear_analysis}. The main conclusions of our study
can be summarized as follows :

\begin{itemize}
\item The linear analysis results are confirmed by our numerical
  simulations: thermal and lepton number diffusion alleviates the
    stabilising effect of buoyancy, and allows MRI growth in regions
  of parameter space where the MRI would be stabilised in the absence
  of diffusion \citep{menou04,masada06, masada07}. The growth rate of the MRI
  is then largely controlled by the thermal and lepton number
  diffusion, faster diffusion leading to faster MRI growth.
\item In the non-linear phase of the MRI, stable buoyancy  (\ie the square
    of the Brunt-V\"ais\"al\"a frequency $N^2>0$) favours
  the appearance of large scale coherent flows: vertically varying
  (but horizontally uniform) channel flows as well as radially varying
  (but azimuthally and vertically uniform) zonal flows. These coherent
  flows can amount to a significant fraction of the total flow
  energies. In the case of high
  thermal diffusion, they are quasi-stationary and very stable over
  time, while they show recurrent phases of formation and disruption
  in the case of low thermal diffusion.
\item The strength of MHD turbulence (characterized by the magnetic
  energy or the rate at which energy is injected into turbulence)
  increases significantly in the regime of unstable buoyancy {\bf (\ie
    $N^2>0$)}, but it decreases only mildly if the flow
    is buoyantly stable. This general result is
  robust, \ie it holds even when the thermal and lepton number
  diffusion is varied by a factor ten. The precise variation of turbulence
    strength with $N^2$ does, however, depend on the thermal diffusion. For high
  diffusion, the magnetic energy decreases monotonically when the
    flow becomes more and more stable against buoyancy (following the
  same qualitative trend as the growth rate). In the case of lower
  diffusion, however, the magnetic energy increases for flows
    that are very stable against buoyancy, an opposite trend to the
  growth rate which decreases dramatically. This surprising behaviour
  is observed in the regime where recurrent channel flows appear, and
  we speculate that it might be related to the effect of buoyancy on
  parasitic instabilities growing on MRI channel modes.
  \item Importantly, we find that the amplitude at which the first exponential growth of the MRI is terminated is not necessarily correlated with the strength of the turbulent phase that follows. Indeed, in the realistic case of high thermal diffusion, the termination amplitude of the MRI increases with $N^2$ while the strength of turbulence decreases. This should be kept in mind when interpreting the results of simulations focusing on the MRI termination \citep{obergaulinger09}, and highlights the importance of studying the fully turbulent phase of the MRI.
\item Other properties of MRI turbulence are strongly influenced by
  buoyancy. In particular, the ratios of azimuthal to radial magnetic
  field and of azimuthal to radial velocity increase dramatically in
  the regime of buoyantly stable flows, where the azimuthal components
  of the magnetic field and of the velocity therefore dominate the
  magnetic and kinetic energy. In the regime of buoyantly unstable
  flows, the ratio of kinetic to magnetic energy increases while the
  Reynolds stress changes sign and becomes negative. All of these
  trends are at least qualitatively reproduced by the properties of
  linear unstable modes.
\end{itemize}

The magnetic field strength estimated from the volume and time
averaged magnetic energy in the non-linear phase of our simulations
ranges from $3.5\times10^{14}\,{\rm G}$ in the case of stable buoyancy
with high thermal diffusion to $10^{15}\,{\rm G}$ in the unstable
buoyancy regime. Such magnetic field strengths lie the range estimated for
magnetars, and they are a few tens to almost a hundred times stronger
than the initial magnetic field. This provides support to the ability
of the MRI to amplify the initial magnetic field to magnetar
strength. Furthermore, if the MRI is at the origin of a magnetar's
magnetic field, our results suggest that the azimuthal magnetic field
would be significantly stronger than the poloidal one (in stably
stratified regions of the PNS). This could have interesting consequences for magnetars. The observation of magnetar activity from neutron stars with low dipole magnetic field \citep{rea12,rea13,rea14} indicates an internal magnetic field much stronger than its dipolar component. Furthermore, the recent detection of a phase modulation of the pulsed emission of two magnetars has been interpreted as free precession arising from a deformation of the magnetar due to a very strong internal toroidal field \citep{makishima14,makishima15}. The generation by the MRI of an azimuthal magnetic field much stronger than its poloidal component could be important to explain these observations.

It has been suggested that the development of the MRI could impact the
dynamics of core collapse supernovae explosions \citep{akiyama03} by
converting shear energy into thermal energy \citep{thompson05} or by
generating large scale magnetic fields that could then drive powerful
jet-like explosions \citep[e.g.][]{moiseenko06, shibata06, burrows07b,
  dessart08, takiwaki09, takiwaki11}. In the former case, the effect
of the MRI was parametrized by an effective turbulent viscosity
estimated to reach values up to a few times
$10^{13}\,{\rm cm^2\,s^{-1}}$ \citep[see Fig.~5 of][]{thompson05}.
The effective turbulent viscosity can be estimated in our simulations
from the angular momentum flux. We found values ranging from
$\sim5\times 10^{10}\,{\rm cm^2\,s^{-1}}$ in the stable buoyancy
regime with high thermal diffusion to
$\sim 5\times10^{11}\,{\rm cm^2\,s^{-1}}$ in the unstable buoyancy
regime. These values are two to three orders of magnitude smaller than
those used by \citet{thompson05}, and they would probably not provide
enough viscous heating to affect the explosion dynamics. Concerning
the second case, the generation of a large scale magnetic field by the
MRI cannot be studied in a local model such as ours, and has so far
never been demonstrated. Though we could not reach definitive
conclusions on this matter, our simulations show that stable buoyancy
favours the formation of coherent magnetic structures on the largest
scales allowed in our local model. The magnetic field strength we
obtained is of the same order of magnitude as that needed to launch
jet-driven explosions ($\sim 10^{15}\,{\rm G}$). In order to launch a jet, such a magnetic field strength is, however, needed near the surface of the PNS, while our simulation domain is located deep inside the PNS. Further numerical simulations describing the outer parts of the PNS will therefore be needed to address this question.

We stress that our quantitative conclusions (e.g. strength of the
magnetic field, angular momentum flux) should not be considered as
definitive answers due to several limitations of our study. Firstly,
as discussed in Section~\ref{sec:box_size}, the results may depend on
the size of the computational domain (a limitation shared by all local
and semi-global simulations). This size dependence in the presence of
buoyancy is unknown so far and should definitely be investigated in
future studies. Secondly, the magnetic Prandtl number assumed in our
simulations due to numerical constraints ($P_m=4$) is much smaller
than its realistic value ($P_m\sim 10^{13}$). In the absence of
buoyancy, MRI turbulence is known to be very sensitive to the magnetic
Prandtl number, the strength of turbulence strongly increasing with
$P_m$ \citep{lesur07, fromang07b, longaretti10}. If this trend were to
hold in the presence of buoyancy, our results should be considered as
lower bounds on the magnetic field strength and the angular momentum
flux.

Many questions remain open on the MRI in the presence of buoyancy. As
already mentioned, it will be crucial to determine the dependence of
our results on the box size and the magnetic Prandtl number. The role
of the strength and geometry of the initial magnetic field (both of
which are very uncertain) should also be explored. According to models
of stellar evolution including magnetic effects approximately, the
initial magnetic field might be dominated by its azimuthal component
\citep{heger05}. Is an efficient amplification of the magnetic field
possible, if the initial magnetic field is azimuthal rather than
poloidal? How do the results depend on the strength of the initial
magnetic field? Further numerical simulations in the framework laid
out in this paper could answer such questions. Note also, that this study was restricted to the equatorial plane of the PNS, where the stratification is perpendicular to the rotation axis. It would be interesting to investigate the impact of buoyancy at other latitudes, thereby allowing for a range of angles between the radial stratification and the rotation axis.

Finally, this study was restricted to the inner region of the PNS
where neutrinos are in the diffusive regime. \citet{guilet15} showed
that in the outer parts of the PNS, but still below the
neutrinosphere, neutrinos are in the non-diffusive regime on
length scales at which the MRI grows. The influence of buoyancy on the
MRI in this regime is unknown, and should be studied using both linear
analysis and numerical simulations.

\section*{Acknowledgments}
We thank Geoffroy Lesur for providing us with the snoopy code and for useful discussions. We thank Henrik Latter, Vincent Pratt, Yudai Suwa and Thomas Janka for useful discussions. JG acknowledges support from the
Max-Planck-Princeton Center for Plasma Physics.

\appendix

\section{MHD linear analysis}
\label{sec:linear_analysis}
In this appendix we give an analytical description of the MRI modes
growing in the setup described in Section~\ref{sec:setup}. The
dispersion relation obtained in Section~\ref{sec:dispersion_MHD} is
equivalent to that obtained by \citet{menou04} and \citet{masada07},
but restricted to the particular setup considered in this paper, \ie
for a purely vertical magnetic field, the description of buoyancy
effects by a single buoyancy variable (relevant to the particular case
where the thermal and lepton diffusivities are equal), and considering
only the dynamics in the equatorial plane. We repeat the derivation of
the respective dispersion relation below, because for our setup the
derivation is easier to follow, and because intermediate results
(relations between the different velocity and magnetic field
components of the modes) are subsequently used in
Sections~\ref{sec:linear_stress} and \ref{sec:linear_energy}.

\subsection{dispersion relation}
\label{sec:dispersion_MHD}
For simplicity, the analysis is restricted to MRI channel modes whose
wave vectors are purely vertical (\ie having only a $z$ component),
because these are the fastest growing modes in the presence of a
purely vertical background magnetic field. Perturbations from the
stationary background flow are assumed to have a time and space
dependence $\delta A = Re\lbrack \delta\tilde{A}\, e^{\sigma t + ikz
}\rbrack$ for $A \in \{u_x, u_y, u_z, B_x, B_y, B_z, \theta\}$, where
  $\sigma$, $k$, and $\delta \tilde{A}$ are the growth rate, the
  vertical wavevector, and the complex amplitude of the mode,
  respectively. For conciseness, we drop the $\tilde{}$ symbol when
referring to the complex amplitudes in the rest of this section. The
equations governing the time evolution of such perturbations are
obtained from Eqs.~(\ref{eq:base1})--(\ref{eq:base5}). They read
\begin{eqnarray}
  \label{eq:pert1}
  \sigma \delta u_x &=& \frac{B_z}{4\pi\rho_0}ik\delta B_x  + 2 \Omega \delta u_y
                        - N^2\delta\theta - k^2 \nu \delta u_x \\
  \label{eq:pert2}
  \sigma \delta u_y -\mathrm{S} \delta u_x &=& \frac{B_z}{4\pi\rho_0}ik\delta B_y
                                               -2 \Omega \delta u_x - k^2 \nu \delta u_y\\
  \label{eq:pert3}
  \delta u_z &=& 0 \\
  \label{eq:pert4}
  \sigma \delta B_x&=& B_z ik \delta u_x - k^2\eta\delta B_x,\\
  \label{eq:pert5}
  \sigma \delta B_y
                    &=& B_z ik \delta u_y - S \delta B_x - k^2\eta\delta B_y,\\
  \label{eq:pert6}
  \delta B_z &=&0,\\
  \label{eq:pert7}
  \sigma \delta \theta  &=& \delta u_x - k^2 \chi \delta \theta,
\end{eqnarray}
where $\mathrm{S}=q\Omega$ is the shear rate with $q$ given by equation~(\ref{eq:q}).
Note that these equations are valid for any amplitude of the
perturbations, as no linearization had to be done to obtain them. The
channel mode solutions that we obtain below are therefore non-linear
solutions, just like the classical MRI channel modes in the
incompressible limit \citep{goodman94}. Next, we define
\begin{equation}
  \sigma_\nu \equiv \sigma + k^2 \nu,
\end{equation}
\begin{equation}
  \sigma_\eta \equiv \sigma + k^2 \eta,
\end{equation}
\begin{equation}
  \sigma_\chi \equiv \sigma + k^2 \chi,
\end{equation}
and rewrite Eqs.~(\ref{eq:pert1}), (\ref{eq:pert2}), (\ref{eq:pert4}),
(\ref{eq:pert5}), and (\ref{eq:pert7}) as
\begin{eqnarray}
  \label{eq:pert1bis}
  \sigma_\nu \delta u_x &=& \frac{B_z}{4\pi\rho_0}ik\delta B_x + 2 \Omega \delta u_y
                            - N^2\delta\theta, \\
  \label{eq:pert2bis}
  \sigma_\nu \delta u_y &=&  \frac{B_z}{4\pi\rho_0}ik\delta B_y
                            -(2 \Omega-\mathrm{S}) \delta u_x,\\
  \label{eq:pert4bis}
  \sigma_\eta \delta B_x &=& B_z ik \delta u_x,\\
  \label{eq:pert5bis}
  \sigma_\eta \delta B_y &=& B_z ik \delta u_y - S \delta B_x,\\
  \label{eq:pert7bis}
  \sigma_\chi \delta \theta  &=& \delta u_x.
\end{eqnarray}

Using Eqs.~(\ref{eq:pert1bis}) and
(\ref{eq:pert4bis})--(\ref{eq:pert7bis}), one can express all
non-vanishing variables as a function of the radial velocity amplitude
\begin{equation}
  \delta u_y  = \frac{1}{2 \Omega } \left( \sigma_\nu + \frac{N^2}{\sigma_\chi}
    + \frac{k^2v_A^2}{\sigma_\eta} \right) \delta u_x,
  \label{eq:uy_ux}
\end{equation}
\begin{equation}
  \delta B_x = B_z \frac{ik\delta u_x}{\sigma_\eta},
  \label{eq:Bx_ux}
\end{equation}
\begin{equation}
  \delta B_y = \left( \sigma_\nu + \frac{N^2}{\sigma_\chi} + \frac{k^2v_A^2
      - 2\Omega \mathrm{S}}{\sigma_\eta} \right) \frac{B_z}{\sigma_\eta}
  \frac{ik \delta u_x}{2 \Omega},
  \label{eq:By_ux}
\end{equation}
\begin{equation}
 \delta \theta  = \frac{\delta u_x}{\sigma_\chi}.
  \label{eq:theta_ux}
\end{equation}
Using equation~(\ref{eq:pert2bis}), we then obtain the dispersion relation
\begin{eqnarray}
  \sigma_\chi\left\lbrack\left(\sigma_\nu\sigma_\eta + k^2v_A^2\right)^2
  + \kappa^2 \left(\sigma_\eta^2 + k^2v_A^2 \right)
  - 4\Omega^2k^2v_A^2  \right\rbrack   \nonumber \\
  + N^2\sigma_\eta\left(\sigma_\nu\sigma_\eta + k^2v_A^2\right)  = 0,
  \label{eq:dispersion_MHD}
\end{eqnarray}
where $\kappa$ is the epicyclic frequency defined in equation~(\ref{eq:def_kappa}). 
The more general equation~(31) of \citet{masada07} reduces to our 
dispersion relation in the case of a purely vertical magnetic field
and wavevector, provided the thermal and lepton number diffusivities
are equal (in which case our Brunt-V\"ais\"al\"a frequency should be
identified with the sum of two frequencies of \citet{masada07} that
are related to the entropy and lepton number gradient, respectively)
or either the lepton number gradient or the entropy gradient vanishes.
Equation~(\ref{eq:dispersion_MHD}) is also equivalent to equation~(13) of
\citet{menou04}, when applied in the equatorial plane to a
purely vertical magnetic field and wavevector.

The dispersion relation is a fifth order polynomial in $\sigma$, which
can be written in the form
\begin{equation}
  a_5\sigma^5 + a_4\sigma^4 + a_3\sigma^3 + a_2\sigma^2 + a_1\sigma + a_0 = 0,
\end{equation}
with
\begin{eqnarray}
  a_5&=& 1, \\
  a_4&=&  k^2\left(\chi + 2\nu + 2\eta \right), \\
  a_3&=& N^2 + \kappa^2 + 2k^2v_A^2 \nonumber \\
     & & + k^4\left\lbrack2\chi(\eta+\nu) + 4\eta\nu + \nu^2 + \eta^2\right\rbrack,  \\
  a_2&=& N^2k^2(\nu + 2\eta) + \kappa^2k^2(\chi + 2\eta)  \nonumber \\
     & & + 2k^4v_A^2(\nu + \eta + \chi) \nonumber \\
     & & + k^6\left\lbrack \chi(\nu^2 + \eta^2 + 4\eta\nu) + 2\nu\eta(\eta+\nu)
         \right\rbrack,  \\
  a_1&=& k^8\left\lbrack \eta^2\nu^2 + 2\chi(\eta\nu^2 + \nu\eta^2) \right\rbrack
         \nonumber \\
     & & + N^2k^4(\eta^2 + 2\eta\nu) + \kappa^2k^4(\eta^2 + 2\eta\chi) \nonumber \\
     & & + k^2v_A^2\left\lbrack 2k^4(\nu\eta + \nu\chi + \eta\chi)+ N^2 \right.
         \nonumber \\
     & & \phantom{+k^2v_A^2\,}\left.
         +k^2v_A^2 + \kappa^2 - 4\Omega^2 \right\rbrack, \\
  a_0&=& k^{10}\chi\eta^2\nu^2 + N^2k^6\nu\eta^2 + \kappa^2k^6\chi\eta^2 \nonumber \\
     & & + k^2v_A^2\left\lbrack 2k^6\nu\eta\chi + N^2k^2\eta \right.
         \nonumber \\
     & & \phantom{+k^2v_A^2\,}\left.
         +(k^2v_A^2 + \kappa^2 - 4\Omega^2)k^2\chi \right\rbrack. 
\end{eqnarray}
In order to obtain the growth rates, this fifth order polynomial is
solved numerically using a Laguerre method.

\subsection{Reynolds and Maxwell stresses}
\label{sec:linear_stress}
The vertical average of the Maxwell and Reynolds stresses associated
with a channel mode can be computed using the complex amplitudes of the
velocity and magnetic field perturbations. The vertical average of the product of
two quantities $f$ and $g$ is obtained from their complex amplitudes
$\tilde{f}$ and $\tilde{g}$ through
\begin{equation}
  \left\langle fg \right\rangle = \frac{1}{L_z}\int_{-L_z/2}^{L_z/2} fg \dd z
  = \frac{1}{2}{\rm Re}(\tilde{f}\tilde{g}^*),
\end{equation}
where $\tilde{g}^*$ is the complex conjugate of $\tilde{g}$. Here, we
have used the relation $Re(z_1)Re(z_2) = (z_1+z_1^*)(z_2+z_2^*)/4 =
\left[Re(z_1z_2^*) + Re(z_1z_2)\right]/2$, for two complex numbers
$z_1$ and $z_2$, and the fact that the vertical average of
$Re\left(\tilde{f}\tilde{g}\,e^{2\sigma t + 2ikz }\right)$ vanishes
(for non vanishing $k$).

Using equation~(\ref{eq:uy_ux}), the vertical average of the Reynolds
stress reads
\begin{eqnarray}
  R_{xy}      & =    & Re\left( \sigma_\nu + \frac{N^2}{\sigma_\chi}
                  + \frac{k^2v_A^2}{\sigma_\eta} \right)
                  \rho_0\frac{|\delta u_x|^2}{4 \Omega }
\end{eqnarray}
For an unstable mode ($\sigma>0$) and $N^2 \geqslant 0$, the Reynolds
stress is always positive, which is usually the case for MRI channel
modes and turbulence \citep[e.g.][]{pessah06a}. When $N^2 \leqslant 0$
on the other hand, the buoyancy can potentially make the Reynolds
stress take negative values.

Using Eqs.~(\ref{eq:Bx_ux}) and (\ref{eq:By_ux}), we obtain the
vertical average of the Maxwell stress
\begin{eqnarray}
  M_{xy}  &=     & -Re\left[\frac{k^2v_A^2}{\sigma_\eta^2}
                  \left( \sigma_\nu + \frac{N^2}{\sigma_\chi} +
                  \frac{k^2v_A^2 - 2\Omega S}{\sigma_\eta}
                  \right)\right] \nonumber \\
         &      & \times \rho_0\frac{|\delta u_x|^2}{4 \Omega }.
\end{eqnarray}
Using the dispersion relation, this expression can be rewritten as
\begin{equation}
  M_{xy} =  Re\left\lbrack \frac{\sigma_\nu}{\sigma_\eta}
    \left( \sigma_\nu + \frac{N^2}{\sigma_\chi} + \frac{k^2v_A^2}{\sigma_\eta}
    \right) + \frac{\kappa^2}{\sigma_\eta}
  \right\rbrack \rho_0\frac{| \delta u_x |^2}{4 \Omega }.
\end{equation}
The same remarks as for the Reynolds stress apply for the Maxwell
stress. However, note that the term proportional to $\kappa^2$ makes
it less likely for the Maxwell stress to become negative,
because it would require more negative values of $N^2$ than for the
Reynolds stress.

\subsection{Kinetic and magnetic energies}
\label{sec:linear_energy}
In a similar way as for the Reynolds and Maxwell stresses, one can
obtain the vertically averaged kinetic and magnetic energies
\begin{equation}
E_{K}=  Re\left\lbrack 1 + \left(\frac{\sigma_\nu +  N^2/\sigma_\chi +
        k^2v_A^2/\sigma_\eta}{4 \Omega}\right)^2 \right\rbrack \rho_0| \delta u_x |^2,
\end{equation}
\begin{eqnarray}
  E_{M}      &=     & Re\left\lbrack 1 +
                 \left( \frac{ \sigma_\nu + N^2/\sigma_\chi
                 + (k^2v_A^2 - 2\Omega S)/\sigma_\eta}{4 \Omega}
                 \right)^2 \right\rbrack \nonumber \\
        &       &\times \frac{k^2v_A^2}{Re(\sigma_\eta^2)}\rho_0| \delta u_x |^2.
\end{eqnarray}

\section{Hydrodynamical linear analysis}
\label{sec:hydro_linear_analysis}

The dispersion relation in a hydrodynamical flow can be obtained from
the MHD dispersion relation by setting the Alfv\'en velocity to zero
in equation~(\ref{eq:dispersion_MHD})
\begin{equation}
  \sigma_\chi \left(\sigma_\nu^2  + \kappa^2 \right) + N^2\sigma_\nu = 0.
  \label{eq:dispersion_hydro}
\end{equation}
This third order polynomial in $\sigma$ may be written also in the form
\begin{equation}
  a_3\sigma^3 + a_2\sigma^2 + a_1\sigma + a_0 = 0,
\end{equation}
with
\begin{eqnarray}
  a_3&=& 1, \\
  a_2&=& k^2\chi + 2k^2\nu, \\
  a_1&=& k^4\nu^2 + 2k^4\nu\chi + \kappa^2 + N^2, \\
  a_0&=& k^2\chi\left( k^4\nu^2 + \kappa^2 \right) + k^2\nu N^2.  
\end{eqnarray}
In order to obtain the growth rates shown in Fig.~\ref{fig:growth_rate},
we solve numerically this third order polynomial using a Laguerre method.

\begin{figure*}
  \centering
  \includegraphics[width=\columnwidth]{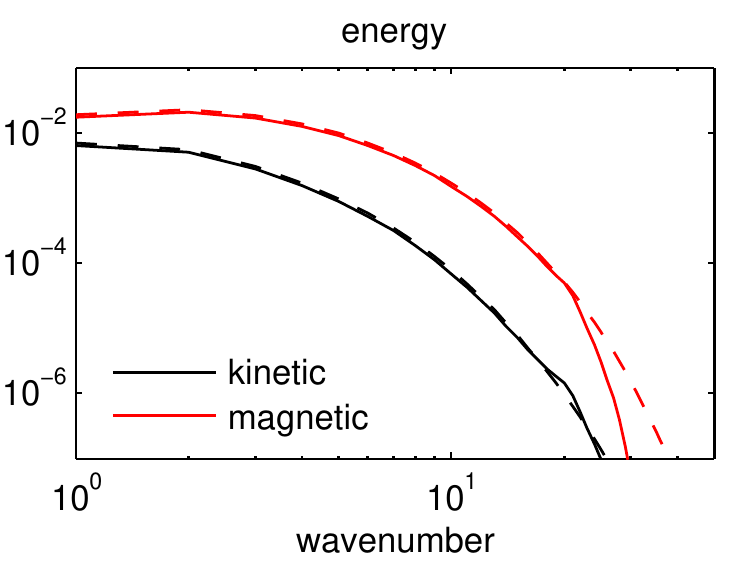}
  \includegraphics[width=\columnwidth]{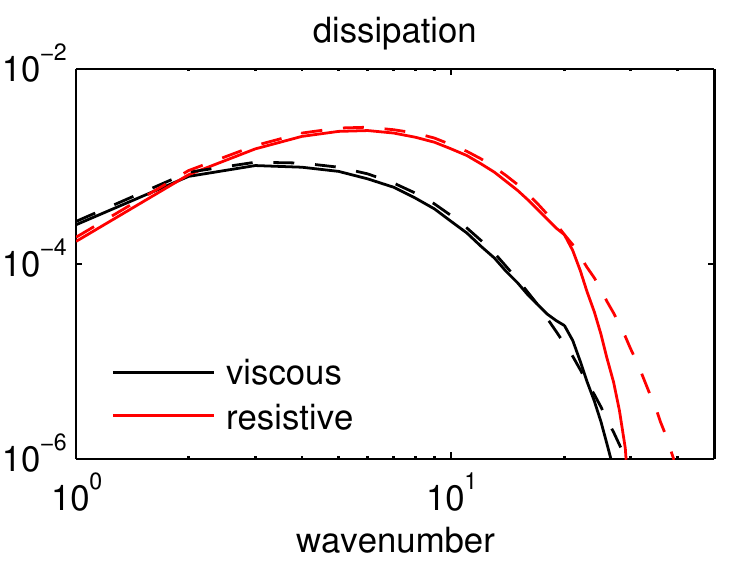}
  \caption{Convergence test for simulations with $N^2=0$ and
    $P_e=100$. Results for the standard grid resolution
    $[n_x,n_y,n_z]=[256,128,64]$ are shown by solid lines, while
    results obtained at a twice higher resolution run are shown by
    dashed lines. Left panel: spectra of the kinetic energy (black
    lines) and magnetic energy (red lines). Right panel: spectra of the
    viscous dissipation rate (black lines) and the resistive
    dissipation rate (red lines). The spectra are averaged over
    spherical shells in Fourier space following \citet{lesur11}, and
    over time for $200\,{\rm ms} \leq t \leq 628\,{\rm ms}$. The
    wavenumber is normalized to $2\pi/L_z$.}
  \label{fig:conv_spectrum}%
\end{figure*}

An interesting approximate analytical solution of the dispersion
relation can be obtained in the inviscid limit by performing an
asymptotic expansion for $N^2 \ll \kappa^2$. The zeroth order solution
is a simple epicyclic oscillation of frequency $\kappa$ and vanishing
growth rate. Performing the expansion up to first order provides the
complex growth rate for small $N^2$
\begin{equation}
  \sigma \simeq i \kappa \left\lbrack1 + \frac{N^2}{2(k^4\chi^2 + \kappa^2)}
  \right\rbrack - \frac{N^2 k^2\chi}{2(k^4\chi^2 + \kappa^2)}
  \label{eq:sigma_hydro}
\end{equation}
where the imaginary part of the growth rate is the oscillation
frequency and the real part is the growth rate. The fastest growing
mode has the following wavenumber and complex growth rate
\begin{equation}
  k_{\rm max} \simeq \sqrt{\kappa/\chi},
  \label{eq:k_max_hydro}
\end{equation}
\begin{equation}
  \sigma \simeq i \kappa \left\lbrack1 + \frac{N^2}{4\kappa^2}
  \right\rbrack
  - \frac{N^2}{4\kappa}.
  \label{eq:sigma_max_hydro}
\end{equation}
We found this analytical solution to be a rather good approximation in
the inviscid limit even for $N^2$ of the order of $\kappa^2$. The
maximum growth rate that we obtain is the same as the one derived by
\citet{lyra14} for $N^2 \ll \kappa^2$ and a purely vertical wavevector
(although these authors considered a thermal relaxation for a given
thermal time rather than the diffusion approximation that we used here
to describe length scales longer than the neutrino mean free path).

\section{Convergence test}
\label{sec:convergence}

We performed a convergence test to check whether the spatial
resolution employed in our simulations was sufficient to resolve the
dissipation scales, and hence whether the results are unaffected by
the limited grid resolution. The spectra of the energy and dissipation
rates are shown in Fig.~\ref{fig:conv_spectrum} for $N^2=0$ and
$P_e=100$, comparing a run performed with our standard resolution
$[n_x,n_y,n_z]=[256,128,64]$ and a run performed at a twice higher
resolution. The energy and dissipation spectra are in good agreement
for $kL_z/2\pi < 20$,  \ie in a regime where most of the energy is
  located and most of the dissipation occurs. The dissipation rate for
$kL_z/2\pi > 20$ (where the spectra start to differ) is at least an
order of magnitude smaller than its peak value. Thus, these unresolved
scales have little impact on the dynamics.

When varying the Brunt-V\"ais\"al\"a frequency, we observed for
$N^2<0$ the peak of the dissipation rate shifted to smaller
  scales due to the enhanced turbulent activity. We therefore increased
the resolution in order to ensure that the peak of viscous, resistive
and thermal dissipation is well resolved.

\bibliography{supernovae}

\bsp
\label{lastpage}

\end{document}